\documentclass{aa}  

\usepackage{graphicx}
\usepackage{txfonts}
\usepackage{soul}
\usepackage{rotating} 

\usepackage{hyperref}
\hypersetup{
    colorlinks=true,
    linkcolor=blue,
    filecolor=magenta,      
    urlcolor=cyan,
    citecolor=blue,
    pdfpagemode=UseThumbs,
    }

\usepackage{makecell}
\usepackage{float}
\usepackage{natbib}
\usepackage{xcolor}
\usepackage[normalem]{ulem}
\usepackage{placeins}
\usepackage{stfloats}
\bibpunct{(}{)}{;}{a}{}{,}
\usepackage{afterpage}
\usepackage{amsmath}
\usepackage{yfonts}
\usepackage{marginnote} 

\defcitealias{2023paperI}{Paper I}
\defcitealias{2024paperII}{Paper II}

\newcommand{\obd}{18OB}
\newcommand{\obc}{4OB}
\newcommand{\grey}{grey}

\begin{document} 
   \title{The effect of local magnetic fields in quiet regions of stellar atmospheres simulated with \texttt{MANCHA}} 

   \author{A. Perdomo García
          \inst{1}
          \and
          N. Vitas\inst{2}\fnmsep\inst{3}
          \and 
          E. Khomenko
          \inst{2}\fnmsep\inst{3}
          \and 
          M. Collados
          \inst{2}\fnmsep\inst{3}
          }

   \institute{Max-Planck Institut für Astronomie, 
              69117 Heidelberg, Germany
              \\
              \email{andrperdomo@gmail.com}
        \and
              Instituto de Astrofísica de Canarias,
              38200 La Laguna, Tenerife, Spain  
         \and
             Departamento de Astrofísica de la Universidad de La Laguna,
             38200 La Laguna, Tenerife, Spain
             }

   \date{Received Month day, year; accepted Month day, year}

  \abstract
   {In the last years, new methods that allow the measurement of local magnetic fields in photospheres of distant stars have been developed. The first simulations of small-scale dynamo in other stars than the Sun have been produced too. While the nature of the small-scale fields is still under debate, it is an accepted fact that they can be generated by the action of small-scale dynamo in simulations.
    }
   {Our aim is to characterize the effects of the local magnetic fields in quiet regions of stellar atmospheres. }
   {We compute magneto-hydrodynamic and purely hydrodynamic simulations of  G2V, K0V and M2V star. The magnetic simulations are started from the hydrodynamical ones, adding the Biermann battery term in the induction equation to produce a magnetic seed, that is enhanced by the action of the small-scale dynamo. Once the magnetic field is saturated, we compare the simulations with and without magnetic fields and characterize the differences in statistics of velocities, appearance of granulation, and the mean stratification of a number of relevant parameters. These differences are also compared with the deviations produced by different treatments of the opacity in the simulations. }
   {The saturation values of the magnetic fields are $\sim 100$ G for the three stars in their surface, consistent with the recent results for cool stars, and other results for the Sun in the literature. 
   The local magnetic fields have a negligible effect on  the velocities of the plasma or the mean stratifications of the simulated stars. In contrast, they produce changes in the bolometric intensity of the intergranular lanes and the power spectrum at small scales of the temperature and vertical velocity of downflows. Significant differences between the hydrodynamic and magneto-hydrodynamic simulations are also found for the kinetic energy. This difference in energy can be explained by the transformation of kinetic into magnetic energy, which is consistent with the action of the small-scale dynamo.}
   {}


   \maketitle

\section{Introduction} \label{sec:introduction}

    Magnetic fields play an important role in stellar atmospheres by producing stellar variability and activity. As they are ubiquitous in the solar atmosphere and present in a wide range of magnitudes and scales, they are expected to be found in most main sequence cool stars. For solar-type stars, the variability for time-scales larger than a day \citep{2013solanki_unruh} is mainly induced by changes of the magnetic field at a global scale, which is in turn driven by the joint action of convection, turbulence, and differential rotation. 

    There are several possible origins for small scale magnetic fields in quiet regions of the Sun \citep[see the review by][]{2019Bellot_Orozco_review_SSD}, the main theories proposed are the global dynamo described above and a small-scale dynamo (SSD). The SSD, excited by turbulent scales, can be capable of enhancing magnetic seeds up to saturation levels close to those in quiet regions. The global dynamo could contribute to the origin of local fields in different ways, for example, by producing active regions that give flux to smaller scales during their decay or by generating flux in deep layers that is then redistributed in different scales during its emergence. While there is controversy regarding whether SSD can develop in the real Sun due to the necessary Prandtl regime \citep{2004Schekochihin, 2005Schekochihin, 2018Brandenburg, 2023Warnecke_SSD_smallPrandtl}, observations have not been able to support the large-scale dynamo hypothesis, since no changes in the quiet regions are observed along the global magnetic field cycle \citep{2011lites_hinode_ssd, 2013Buehler, 2014lites_hinode_ssd, 2024solanki}. Nevertheless, simulations have shown that, if present, both mechanisms could contribute significantly to the solar fields in quiet regions. The focus of the present work is on the SSD. 
    \citet{2004TrujilloBueno_SSD} used 3D MHD simulations and the Hanle effect to measure for the first time the local magnetic fields in the quiet Sun (with average strength of $\simeq 130$ G). Later authors used both simulations and observations to obtain similar conclusions \citep{2005Khomenko_SSD, 2010Danilovic, 2011Shchukina, 2012_OS_BR}. 
    \citet{2007voegler_Schussler_SSD} simulated amplification of the turbulent field by the action of SSD and demonstrated that the resulting field reaches the same level of magnitude as in the observations.  
    \citet{2008steiner} compared the local magnetic fields from simulations against observations, and showed that different configurations of the boundary conditions and initial magnetic seed produced the same saturated field. This result was further improved by \citet{2014rempel_SSD} who also studied the effect of resolution and boundary conditions. \citet{2015Kitiashvili} showed in detail the growth of the local fields in the simulation from a magnetic seed. \citet{2017khomenko} showed that convection is inherently magnetized since the magnetic seeds needed for SSD may be created even in pure hydrodynamical models by the action of the Biermann battery. Recently, it has been demonstrated in numerical simulations that SSD operates on the surfaces of other cool stars too \citep{2022bhatia_ssd, 2023_ssd_muram_II, 2023arXiv_ssd_muram_III, 2024Riva_ssd}. 

    From the observational side, methods to measure the small-scale magnetic fields in other stars than the sun have been developed in the last years. 
    One of the techniques used to measure such fields relies on the variation of the intensity of the lines due to Zeeman effect, rather than the broadening (which can be mixed with other broadening sources, e.g. rotation). The Zeeman line intensification depends solely on the magnetic field parameters and the properties of the particular spectral line studied. High-quality observations are not required to measure Zeeman intensification, although it takes a proper modelling of the magnetic desaturation of the lines. The method has been successfully applied in active M dwarfs using the Ti {\tiny I} multiplet at $9647-9788 \, \AA$ \citep{2017Shulyak, 2019Shulyak, 2017Kochukhov, 2019Kochukhov} and in 15 solar-like stars using the Fe {\tiny I} lines at $5497.5, \, 5501.6, \,\mathrm{and} \, 5506.8 \, \AA$ \citep{2020Kochukhov}. 

    An additional method is presented in \citet{2024Guifang}, where they used   
    one-dimensional (1D) stellar models to constrain the local magnetic fields of the solar analogue KIC 8006161. Aided by asteroseismology, the oscillation spectrum in stellar atmospheres can be used to characterize the waves in the plasma. Waves provide a convenient proxy for magnetic fields, since their propagation speed is modified by their presence. 
    This method was previously used to prove the local magnetic fields of the Sun \citep{2021Li} and to measure the magnetic field in the interior of giants \citep{2022Li, 2023Deheuvels}. 
    Combined with these recent measurements of small-scale magnetic fields compelled by the development of new methods, 3D MHD simulations can be used to know and understand the stellar local magnetic fields of observed stars.

    The action of the SSD in real stars is still an open debate. 
    One important parameter used to determine if SSD operates in a plasma is the magnetic Reynolds number (the ratio of the induction and diffusion of magnetic field in the induction equation) 
    \begin{equation}
        R_\mathrm{m} = \frac{|\vec{\nabla} \times (\vec{\varv} \times \vec{B})|}{|\vec{\nabla} \times (\eta \vec{J})|}
    \end{equation}
    and magnetic Prandtl number
    \begin{equation}
        P_\mathrm{m} = \frac{R_\mathrm{m}}{R_\mathrm{e}}, 
    \end{equation}
    where $\vec{\varv}$ is the velocity vector, $\vec{B}$ is the magnetic field vector, $\vec{J} = (\vec{\nabla} \times \vec{B}) / \mu_0$ is the electric current vector, $\mu_0$ is the magnetic permeability of free space, and $R_\mathrm{e}$ is the Reynolds number (ratio of inertial and viscous forces)
    \begin{equation}
        R_\mathrm{e} = \frac{|\rho (\vec{\varv} \cdot \vec{\nabla}) \cdot \vec{\varv}|}{|\vec{\nabla} \cdot \vec{\mathcal{T}|}}. 
    \end{equation}
    Here, $\rho$ is the mass density and $\vec{\mathcal{T}}$ is the viscous tensor (see e.g. section 15 in \citealp{1987Landau}). 

    The magnetic Prandtl number depends on the numerical scheme employed. It can be either set explicitly in a simulation, or can be determined approximately based on the average diffusive fluxes. When the diffusion is implicit and the treatment of the velocity and magnetic field diffusion is the same, it is found to be  around unity in average in simulations, although values down to $10^{-3}$ are present \citep{2017khomenko}.
    The Prandtl number is thought to be orders of magnitude smaller than unity in real stellar plasmas \citep[down to $10^{-6}$ -- $10^{-4}$, see][]{2002intro_sun_stix} in contrast to the values found in simulations. Although several studies in the literature concluded that the action of the SSD for Prandtl numbers lower than unity was unlikely \citep{2004Schekochihin, 2005Schekochihin, 2018Brandenburg}, with recent experiments and increased resolution it has been demonstrated that it is indeed possible to excite the local dynamo for such values of the Prandtl number \citep{2023Warnecke_SSD_smallPrandtl}. 

    \citet{2022bhatia_ssd} showed for the first time that SSD operates in other main sequence cool stars (ranging between F3V and M0V spectral types). In that work, they characterized the differences in the mean stratification of the convection zone, and pointed out that the four simulated stars reached the same intensity of the saturated magnetic field in the optical surface. \citet{2023_ssd_muram_II} assessed the effect of metallicity (-1, 0 and 0.5) on the small-scale magnetic fields and the mean stratification of a solar-like star. \citet{2023arXiv_ssd_muram_III} did a similar analysis as in \citet{2022bhatia_ssd} for the same stellar types, but focused on the photosphere. They addressed changes in the bolometric intensity for each spectral type, and compared the force balance in a representative magnetic field concentration from the F and M -- type stars. \citet{2024Riva_ssd} characterized the growth rate of the small-scale magnetic fields for a range of cool stars (from K8V to F5V spectral types). They found that the growth rate decreases from the K to the F --type stars, which is explained by the time-scales of the granulation. In all the mentioned works they find bright points in the locations where there are concentration of the local magnetic fields. 
    
    In our previous work \citet[][hereafter \citetalias{2023paperI}]{2023paperI}, we tested several strategies to group the opacity through the binning method \citep{norlund1982}, for four one-dimensional models of main sequence cool stars (F3V, G2V, K0V and M2V spectral types) with solar metallicity. Optimal distributions of the bins were found for each of the four stars. The optimal binned opacities were used in our second paper in this series, \citet[][hereafter \citetalias{2024paperII}]{2024paperII}, which was focused on 3D hydrodynamic (HD) simulations of a G2V, K0V, and M2V star. The simulated stellar atmospheres were consistent with previous works, revealing the same appearance of granulation and statistics of velocity flows as in the literature. Additionally, we studied the differences in the 3D structures of the radiative energy exchange rate $Q$ between the three spectral types. 
    Finally, in the present paper, the HD simulations run in \citetalias{2024paperII} are continued after adding the Biermann battery term in the induction equation to produce a magnetic seed. This seed is then enhanced by the action of SSD for enough stellar time until saturation of the local magnetic field is reached. 
    By comparing the simulations from \citetalias{2024paperII} and the current paper, we focus on the effects that the presence of turbulent magnetic fields in the quiet stellar regions produces on the appearance of the granulation, the statistics of the velocities, and the mean stratification. For that evaluation, we compare the changes produced by the magnetic fields against the ones produced by different approximations to the opacity, studied in 3D domains in \citetalias{2024paperII}. The particular opacity strategies are the Rosseland \grey, four-bins (\obc), and 18-bins (\obd) setups introduced in \citetalias{2024paperII} (see appendix A in the reference).

    In Sect.\ \ref{sec:method} we present the numerical code \texttt{MANCHA} and the setup used to create the simulations in this work. Our results and the comparison of the SSD and HD simulations are presented in Sect.\ \ref{sec:results}. Finally, the conclusions are drawn in Sect.\ \ref{sec:conclusions}.

\section{Methods} \label{sec:method}

    \begin{table*}[t!]
    \caption{Parameters used in the numerical setup.}
    \tabcolsep=0.17cm
    \begin{tabular}{ccccccccccc}
    \hline
      & $N_{\mathrm{x,y}}^2 \times N_{\mathrm{z}}$ & \begin{tabular}[c]{@{}c@{}}$dx \times dy \times dz$\\ $\mathrm{[km]}$\end{tabular} & \begin{tabular}[c]{@{}c@{}}$dt$\\ $\mathrm{[s]}$\end{tabular} &  \begin{tabular}[c]{@{}c@{}}$dt_{\mathrm{s}}$\\ $\mathrm{[s]}$\end{tabular} & \begin{tabular}[c]{@{}c@{}}$\log g$\\ $\mathrm{[cm \, s^{-2}]}$\end{tabular} & \begin{tabular}[c]{@{}c@{}}$T_{\mathrm{eff}}$\\ $\mathrm{[K]}$\end{tabular} & \begin{tabular}[c]{@{}c@{}}$\langle z \rangle_{\mathrm{\tau=1}}$\\ $\mathrm{[km]}$\end{tabular} & \begin{tabular}[c]{@{}c@{}}$L_x \times L_y \times L_z$\\ $\mathrm{[Mm]}$\end{tabular} & \begin{tabular}[c]{@{}c@{}}$t_{\mathrm{tot}}$\\ $\mathrm{[min]}$\end{tabular} \\ \hline
     G2V & $384^2 \times 106$ & $16 \times 16 \times 16$    & 0.05           & 30 & $4.4$ & $5780$ & 1045  &  $6.1\times6.1\times1.69$ & 60 \\ \hline
     K0V & $384^2 \times 130$ & $10 \times 10 \times 7$     & 0.1            & 15 & $4.6$ & $4855$ & 609   & $3.8\times3.8\times0.91$   & 45 \\ \hline
     M2V & $432^2 \times 102$ & $3.5 \times 3.5 \times 3.6$ & $ 0.085$       & 10 & $4.8$ & $3690$ & 280  &   $1.5\times1.5\times0.37$ & 20 \\ \hline
     \end{tabular}
     \label{tab:parameters_sim_1_SSD}
     \tablefoot{Number of cells, size of cells, simulation time step, time step for saving snapshots, gravity, effective temperature, mean geometrical height of $\tau=1$ (with the zero point of the geometrical height at the bottom of the domain), total size of the domain, and total time of stationary convection  
    for the stellar simulations presented in this work.}
    \end{table*}

    \subsection{Simulations of near-surface convection with \texttt{MANCHA}} \label{subsec:convection_mancha}

    We used the \texttt{MANCHA} \citep{2010felipe,2017khomenko,2018khomenko, 2023mancha} code to solve the time-dependent radiative 3D MHD set of equations on a uniform Cartesian grid. The eight primary variables evolved by the code are the mass density $\rho$, the three components of the velocity $\vec{\mathrm{\varv}}$, the three components of the magnetic field $\vec{\mathrm{B}}$, and the energy per unit volume
    \begin{equation}
        e = \frac{1}{2} \rho \varv^2 + \frac{B^2}{2 \mu_0} + e_{\mathrm{int}},
    \end{equation}
    where $e_{\mathrm{int}}$ is the internal energy per unit volume. The system is closed by the equation of state (EOS) and the radiative transfer equation (RTE). The EOS is used to compute the temperature $T$ and the gas pressure $p$ from the density and energy per unit volume, while the RTE is used to compute the radiative energy exchange rate $Q$ that appears as a source term in the energy conservation equation.

    To produce realistic simulations of stellar magneto-convection, we only activate the Biermann battery term in the induction equation, which produces a magnetic field seed that is amplified by the SSD effect. Thermal conduction and other non-ideal effects are possible to account for in \texttt{MANCHA}, but since they are negligible in the photosphere \citep{1956Spintzer, 2014Khomenko, 2018Ballester}, they are not included in the equations. 
    Thus, we use the mass continuity, the momentum conservation, the energy conservation, and the induction equations in the form:

    \begin{equation} \label{eq:mhd1}
        \frac{\partial \rho}{\partial t} + \vec{\nabla} \cdot \left( \rho \vec{\varv} \right) = \left( \frac{\partial \rho}{\partial t} \right)_{\mathrm{diff}}, 
    \end{equation}
    \begin{equation} \label{eq:mhd2}
        \frac{\partial \rho \vec{\varv}}{\partial t} +  \vec{\nabla} \cdot \left[ \rho \left( \vec{\varv} \otimes \vec{\varv} \right) +  \left( p + \frac{\vec{B}^2}{2 \mu_0} \right) \vec{I}  - \frac{\vec{B} \otimes \vec{B}}{\mu_0}  \right] = \rho \vec{g} + \left( \frac{\partial \rho \vec{\varv}}{\partial t} \right)_{\mathrm{diff}}, 
    \end{equation}
    \begin{multline} \label{eq:mhd3.5}
        \frac{\partial e}{\partial t} + \vec{\nabla} \cdot \left[ \vec{\varv} \left( e + p + \frac{\vec{B}^2}{2 \mu_0}\right) - \frac{\vec{B} \left( \vec{\varv} \cdot \vec{B}  \right)}{\mu_0}
        - \frac{\vec{\nabla} p_\mathrm{e} \times \vec{B} }{\textswab{e} n_\mathrm{e} \mu_0} \right] = \\
        \rho \left( \vec{g} \cdot \vec{\varv} \right) + Q + \left( \frac{\partial e}{\partial t} \right)_{\mathrm{diff}}, 
    \end{multline}
    \begin{equation} \label{eq:mhd3}
        \frac{\partial \vec{B}}{\partial t} = \vec{\nabla} \times \left[ \vec{\varv} \times \vec{B}  
        + \frac{\vec{\nabla} p_\mathrm{e}}{\textswab{e} n_\mathrm{e}} 
        \right] + \left( \frac{\partial \vec{B}}{\partial t} \right)_{\mathrm{diff}},
    \end{equation}
    where 
    $\vec{g}$ is surface gravity, $\vec{I}$ is the identity tensor, the scalar product is represented by the symbol ``$\cdot$'', the tensor product by ``$\otimes$'',  and the $\left(\dots\right)_{\mathrm{diff}}$ stands for numerical diffusivity terms. The artificial diffusivity in the induction equation is set to zero. The remaining terms are constructed following \citet[][see e.g. \citetalias{2024paperII} and \citealp{my_thesis}]{voegler_thesis, 2023mancha}, to mimic the physical diffusivities present in the equations. The diffusivity in the mass continuity equation has no physical equivalent, and is present only for stability reasons. To make the simulations more stable, \texttt{MANCHA} also uses a 6th order filtering scheme (described in detail in section 3.4 of \citep{2023mancha}), equivalent to a 6th-order diffusion operator.

    The boundary conditions used in this work to run magneto-convection are those described in \citetalias{2024paperII}, accounting additionally for the magnetic field. The horizontal boundary condition is periodic. 
    The top boundary is closed for vertical mass flux and energy, and imposes stress-free horizontal velocities and strictly vertical magnetic field. 
    The bottom boundary is open for the flow and imposes again vertical magnetic field. The condition of vertical fields ensures that the Poynting flux is zero at the bottom, which facilitates the study of the SSD action in isolation. The bottom boundary also has an implementation of mass and entropy controls after \citet{voegler_thesis}. The mass control ensures that the total mass of the domain remains constant throughout the simulation time. The entropy control imposes homogeneous internal energy and total pressure (sum of gas and magnetic pressure) for the upflows at the bottom boundary. This fixes the entropy at the boundary, which is corrected to account for deviations of the targeted emergent flux of the star. The correction is performed in time-scales close to the Kelvin-Helmholtz time of the domain, scaled by a user-defined factor that can be tuned to optimize the initial transitory phase of the simulation.

    The radiative transfer equation (RTE) is solved using the short characteristics method \citep{shortcharactNLTE} along 24 rays (three per octant). The rays are weighted and distributed using the quadrature A2 from \citet{1963carlson}. The density, opacity, and source function are interpolated linearly in the direction of the rays. Local thermodynamic equilibrium is assumed to solve the RTE. The opacity is approximated by the opacity binning method \citep{nordlund_dravins_1990}, to significantly reduce the number of computations but keeping the accuracy of the bolometric radiative outputs. The opacity tables are constructed using the monochromatic opacities from \texttt{SYNSPEC} \citep{SYNSPEC, synple}, and aided from the opacity distribution functions (ODFs) method \citep{1951labs_ODF, kurucz1993_odf}, see Paper II.

    The EOS is precomputed as a function of the temperature and density, assuming instantaneous chemical equilibrium between 92 atomic and 349 molecular species. Most of the included molecules in the EOS are formed by hydrogen (84 molecules), oxygen (78), fluorine (49), carbon (45), chlorine (44), sulphur (37), and nitrogen (33). We include molecules that play an important role in the  structure and spectrum of stellar atmospheres, such as H$_2$, H$_2$O, CH$_4$, CO, CO$_2$, MgH, CaH, FeH, NaH, OH, SiH, TiO, or VO. Regarding the opacities, a detailed explanation of the atomic and molecular linelists can be found in appendix A from \citetalias{2023paperI}. Both the EOS and opacity tables are computed for solar composition \citep{AG89}.

    \subsection{Numerical setup} \label{subsec:setup_ssd}
     
    The hydrodynamic simulations for the G2V, K0V, and M2V stars described in \citetalias{2024paperII} are continued with the \texttt{MANCHA} code after switching on the Biermann battery term in the induction equation (as in Eq.\ \ref{eq:mhd3}). The simulations are run until saturation of the small-scale magnetic field amplified by the action of the SSD (after around 30-40 granulations cycles, see Fig.\ \ref{fig:saturation_vs_time}), see the discussion in Section 3. Once the saturation is reached, the simulations are continued during the times indicated in the last column of Table\ \ref{tab:parameters_sim_1_SSD}. These time series after saturation of the magnetic field are used as representative of the stationary state and are used to produce the results of the present paper. The numerical setups of the simulations of the present paper are similar (see Table\ \ref{tab:parameters_sim_1_SSD}) to those described in \citetalias{2024paperII}, with few differences, detailed below. 
    
    Here only the 4-bins opacity setup shown in figure A.1 from \citetalias{2024paperII} 
    is used (accordingly, the simulations are continued from the HD simulations run with the same opacity setup). As shown in \citetalias{2024paperII}, we find that 4-bins (\obc\ opacity setup) is sufficiently accurate to study the stellar magneto-convection (and permit faster computations than the 18-bins setup, \obd). All the comparisons in the present paper are done between these HD simulations run with four bins, and the continued SSD simulations run with four bins.

    The values of the mean geometrical height of $\tau=1$ (with the zero point of the geometrical height at the bottom of the domain) in Table\ \ref{tab:parameters_sim_1_SSD} are different from the values in table 1 from \citetalias{2024paperII}. This is due to the non-grey treatment of radiation, since the values shown in \citetalias{2024paperII} (986, 590, and 256 km for the G2V, K0V and M2V, respectively) are from simulations run with Rosseland mean opacity, while the simulations in the present work use four bins. When a simulation is reinitiated after changing the opacity setup, the simulation passes through an additional transitory phase owing to the change in radiative losses, and readjusts the mean height of the optical surface. The mean geometrical height values of $\tau=1$ for the SSD simulations (1045, 609, and 280 km for the G2V, K0V and M2V, respectively) are almost identical to the heights of the corresponding HD runs with four bins (1047, 612, and 281 km for the G2V, K0V, and M2V, respectively).

\section{Results} \label{sec:results}
    
    Below we discuss individual snapshots and mean stratifications of different variables. 
    We show different quantities from the SSD simulations, and compare them against the HD case. In general, the comparison for a certain quantity $x$ is done through the difference between the SSD and HD cases, computed as 
    \begin{equation} \label{eq:diff_delta}
        \Delta^x_{\mathrm{SSD-HD}} = x_{\mathrm{SSD}} - x_{\mathrm{HD}},
    \end{equation}
    where the quantity $x$ is specified in the text or the caption of the corresponding figure. We use the temporal standard deviation $\sigma$ divided by the square root of the number of snapshots $N$ (i.e. standard error of the mean) as a proxy for the errors of the mean stratification of a quantity $x$ for the SSD and HD runs. To avoid overcrowding the plots, instead of showing $\sigma/\sqrt{N}$ for $x$, the propagated error for the difference $\Delta^x_{\mathrm{SSD-HD}}$ is shown, i.e.
    \begin{equation}
        E \left( \Delta^x_{\mathrm{SSD-HD}} \right)=\sqrt{ \left( \sigma_{\mathrm{SSD}}/\sqrt{N_{\mathrm{SSD}}} \right)^2 + \left( \sigma_{\mathrm{HD}}/\sqrt{N_{\mathrm{HD}}} \right)^2 } .
    \end{equation}\label{eq:error_differences}

    To compare the SSD simulations versus the HD simulations, we use the same scales as in \citetalias{2024paperII}, namely, geometrical height, Rosseland optical depth, and number of pressure scale heights \begin{equation} \label{eq:np}
        N_p = \ln \left( p/p_{\tau=1} \right),
    \end{equation}
    where $p_{\tau=1}$ is the pressure at the $\tau=1$ surface and the ratio of pressures $p/p_{\tau=1}$ is calculated in every column of the domain. 
    The geometrical height is mostly used to show the 2D maps of different quantities in the snapshot, and the other two scales are used to compare the mean stratification of the different stars. 
    
    As in \citetalias{2024paperII}, 
    the mean stratification of any variable with geometrical height $z$ is computed as the average over horizontal layers in the original grid of the cubes. In the case of the mean stratification of a quantity versus any other scale (e.g. Rosseland optical depth\footnote{Here we use only the simulations run with 4-bins opacity, but the optical depth scale used in the plots is computed using the Rosseland opacity, as it is convention.}), once the scale is computed in every column of the domain, the quantity is interpolated to the common grid of that scale, and then horizontally averaged. Finally, the results are averaged over one hour of stellar time for the G2V, 45 minutes for the K0V, and 20 minutes for the M2V star. 
    The time length is chosen to ensure that the same number of granules are considered in time and space for the three stars, taking into account the granulation lifetimes in table 4.2 in \citet{beeck_thesis}.  In some cases, the quantities for one particular snapshot are shown in a plot. All figures that refer to a single snapshot use the last snapshot of the time series of the corresponding simulation.

    \subsection{Saturation of the small-scale magnetic field}

\begin{figure}[h!]
        \centering
        \includegraphics[width=8cm]{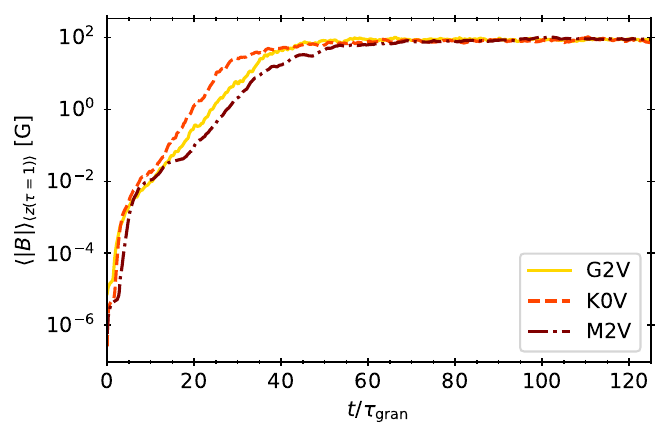}
        \caption{Evolution of the modulus of the magnetic field at the average geometrical height where $\tau = 1$ as a function of the number of granulation cycles $t/\tau_{\mathrm{gran}}$ for the three stars. Here, $\tau_{\mathrm{gran}}$ is the granulation lifetime, equal to 4.18, 3.5, and 1.44 min for the G2V, K0V, and M2V stars, respectively (table 4.2 in \citealp{beeck_thesis}). The G2V star is shown in yellow solid line, K0V in orange dashed, and M2V in brown dot-dashed.}
        \label{fig:saturation_vs_time}
    \end{figure}

    Figure \ref{fig:saturation_vs_time} shows, for the three stars, the temporal evolution of the modulus of the magnetic field $|B|=\sqrt{B_x^2+B_y^2+B_z^2}$ at the average geometrical height where $\tau = 1$ as a function of the 

    \noindent
    number of granulation cycles. The number of granulation cycles is computed as the simulation time $t$ divided by the granulation lifetime $\tau_{\mathrm{gran}}$. The $\tau_{\mathrm{gran}}$ values are taken from table 4.2 in \citealp{beeck_thesis}: 4.18, 3.5, and 1.44 min for the G2V, K0V, and M2V stars, respectively. The initial time in the figure corresponds to the moment in which the simulations are continued after the Biermann battery is switched on in the induction equation. 
    The three stars reach a saturation of the magnetic field of the order of $100$ G after about 30-40 granulation cycles.

    \begin{figure*}[t!]
        \centering
        \includegraphics[width=14cm]{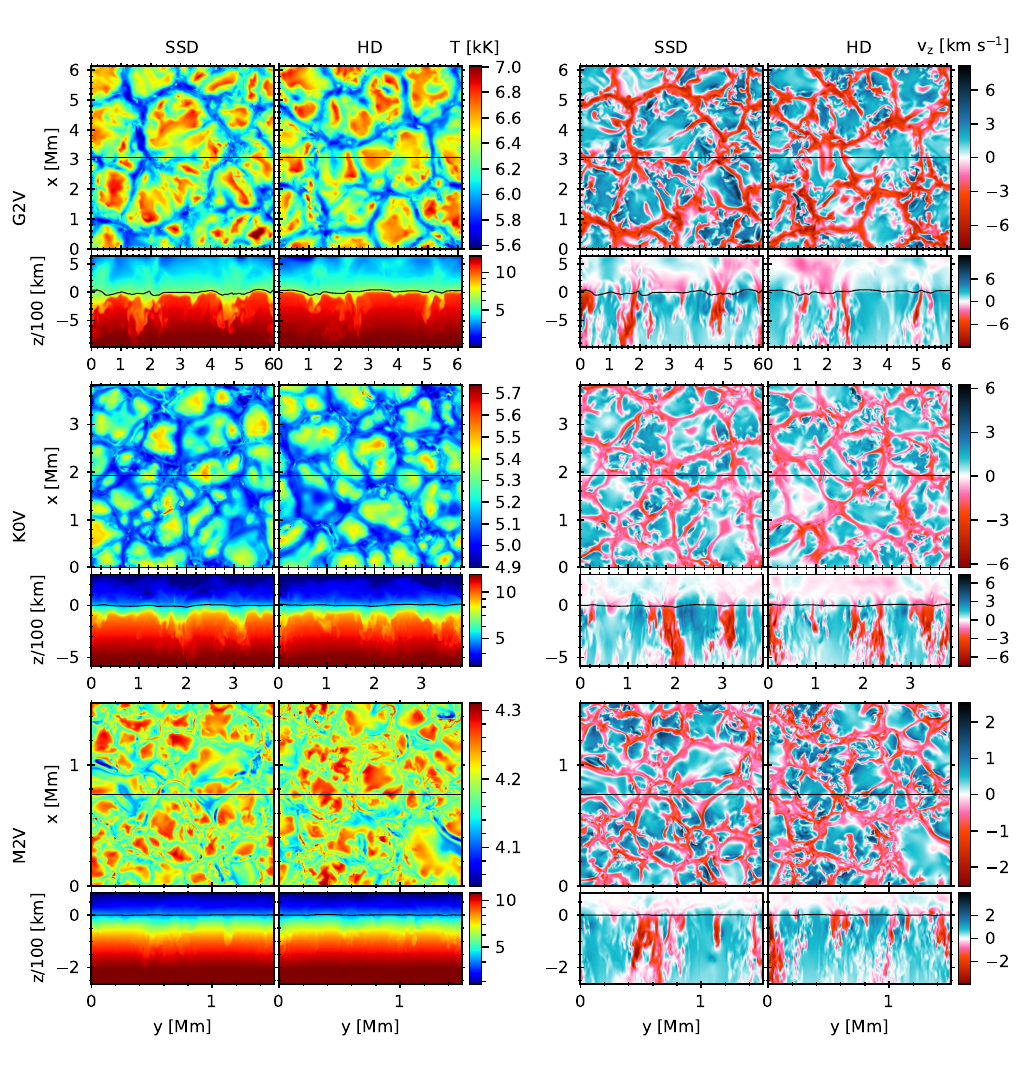}
        \caption{Temperature (left pair of columns), vertical velocity (right pair) of the last snapshot of the SSD (left column in each pair) and HD (right column of each pair) time series of the G2V (top pair of rows), K0V (middle), and M2V (bottom) stars. The values of the quantities shown in the colour-maps are indicated by the corresponding colour-bars (with the same range of values for the SSD and HD panels). The upper panels of each pair of rows shows the surface at $\tau=1$; the lower panel, a vertical cut at the middle of the snapshot. The location of the iso-$\tau$ surface and the vertical cut are shown as black solid curves.}
        \label{fig:2D_quantities_ssd_hd_1}
    \end{figure*}

    \begin{figure*}[t!]
        \centering
        \includegraphics[width=14cm]{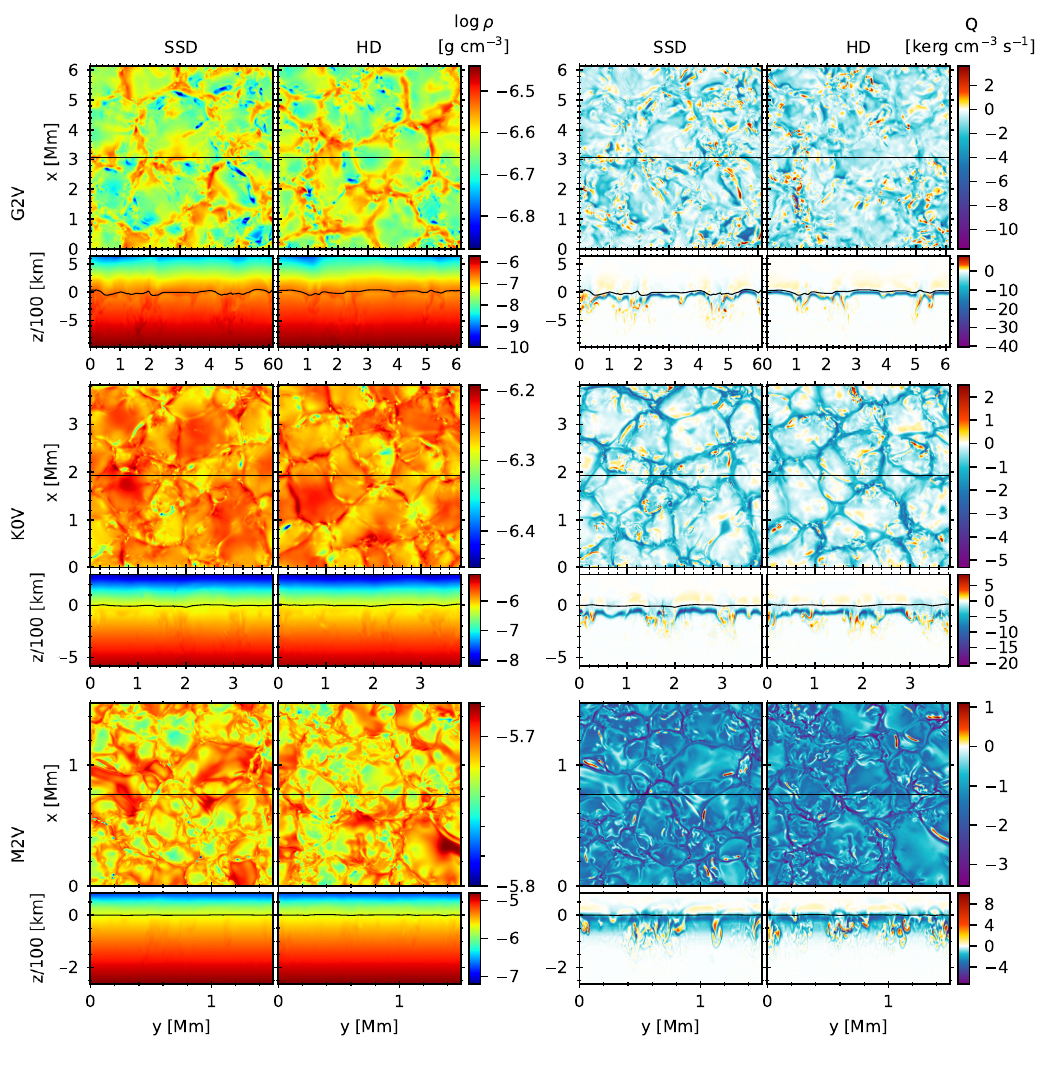}
        \caption{Same as Fig.\ \ref{fig:2D_quantities_ssd_hd_1}, but for logarithm of density (left pair of columns) and $Q$ rate (right).}
        \label{fig:2D_quantities_ssd_hd_2}
    \end{figure*}

    In the temporal evolution of the magnetic field shown in Fig.\ \ref{fig:saturation_vs_time} we can observe the same phases of its amplification detailed in the literature (e.g. see figure 1 in \citealp{2007voegler_Schussler_SSD} or figure 3 in \citealp{2017khomenko}). At the beginning, the effect of the Biermann battery dominates in the induction equation. The magnetic seeds generated by the Biermann battery term are very close for the three stars, of the order of $10^{-6}$ G. Quasi-linear growth of the field is produced in $\lesssim 10$ granulation cycles. After that, the SSD amplifies the field until its saturation. 

    \subsection{Effect on the granulation}

    To compare the appearance of the granulation from the SSD and the HD runs, in Figs. \ref{fig:2D_quantities_ssd_hd_1} and \ref{fig:2D_quantities_ssd_hd_2} we show the temperature, vertical velocity, density, and radiative energy exchange rate $Q$ in the $\tau=1$ surface and a vertical cut in the middle of the last snapshot of the SSD and HD time series for the three stars. To facilitate the comparison, the quantities shown for the SSD and HD panels have the same range of values in the colour-bars, the same as the figure 5 from \citetalias{2024paperII} (except for the panels showing the temperature and density at $\tau=1$, because the density at $\tau=1$ from the K0V and M2V star and the temperature at $\tau=1$ from the M2V star is significantly lower in the grey runs than in the 4-bins runs). A mere visual inspection of the snapshots reveals that the size of the granules is almost equal in the SSD and HD runs. 

    \begin{figure}[t!]
        \centering
        \includegraphics[width=8.8cm]{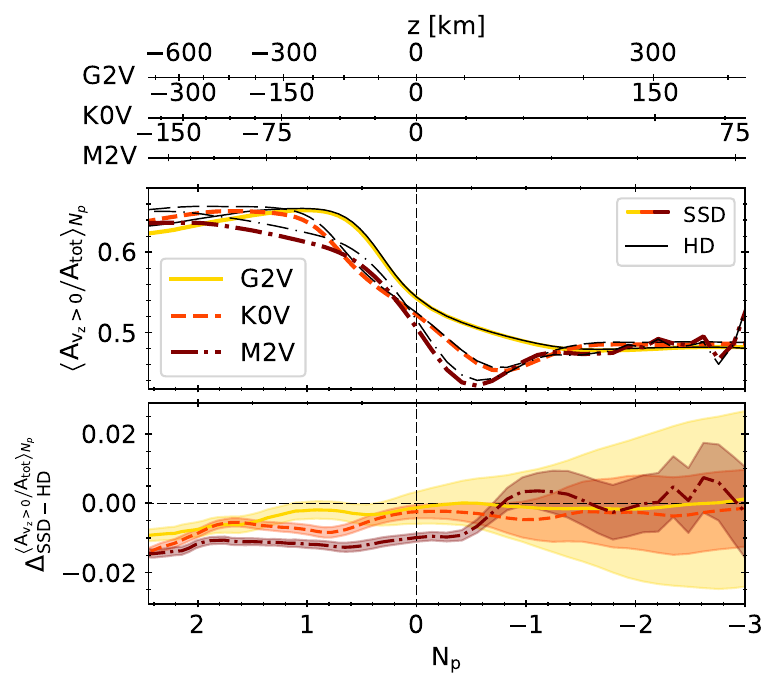}
        \caption{Fractional area covered by upflows of the three stellar simulations averaged over time and surfaces of constant number of pressure scale heights $N_\mathrm{p}$ (Eq. \ref{eq:np}), versus $N_\mathrm{p}$. Top row: the fractional area covered by upflows from the simulations with SSD are shown in yellow solid line for the G2V star, in orange dashed line for the K0V star, and in brown dot-dashed line for the M2V star. The fractional area covered by upflows from the HD simulations is shown in black thin solid, dashed, and dot-dashed lines for the G2V, K0V, and M2V stars, respectively. Bottom row: the difference between the fractional area covered by upflows computed from SSD and HD simulations (Eq.\ \ref{eq:diff_delta}) is shown in yellow solid line for the G2V star, in orange dashed line for the K0V star, and in brown dot-dashed line for the M2V star. The shaded areas show the error of the difference (Eq. \ref{eq:error_differences}). Three top axes: geometrical height for the three stars.}
        \label{fig:granule_size_ssd_vs_hd}
    \end{figure}

    \begin{figure*}[t!]
        \centering
        \includegraphics[width=17.5cm]{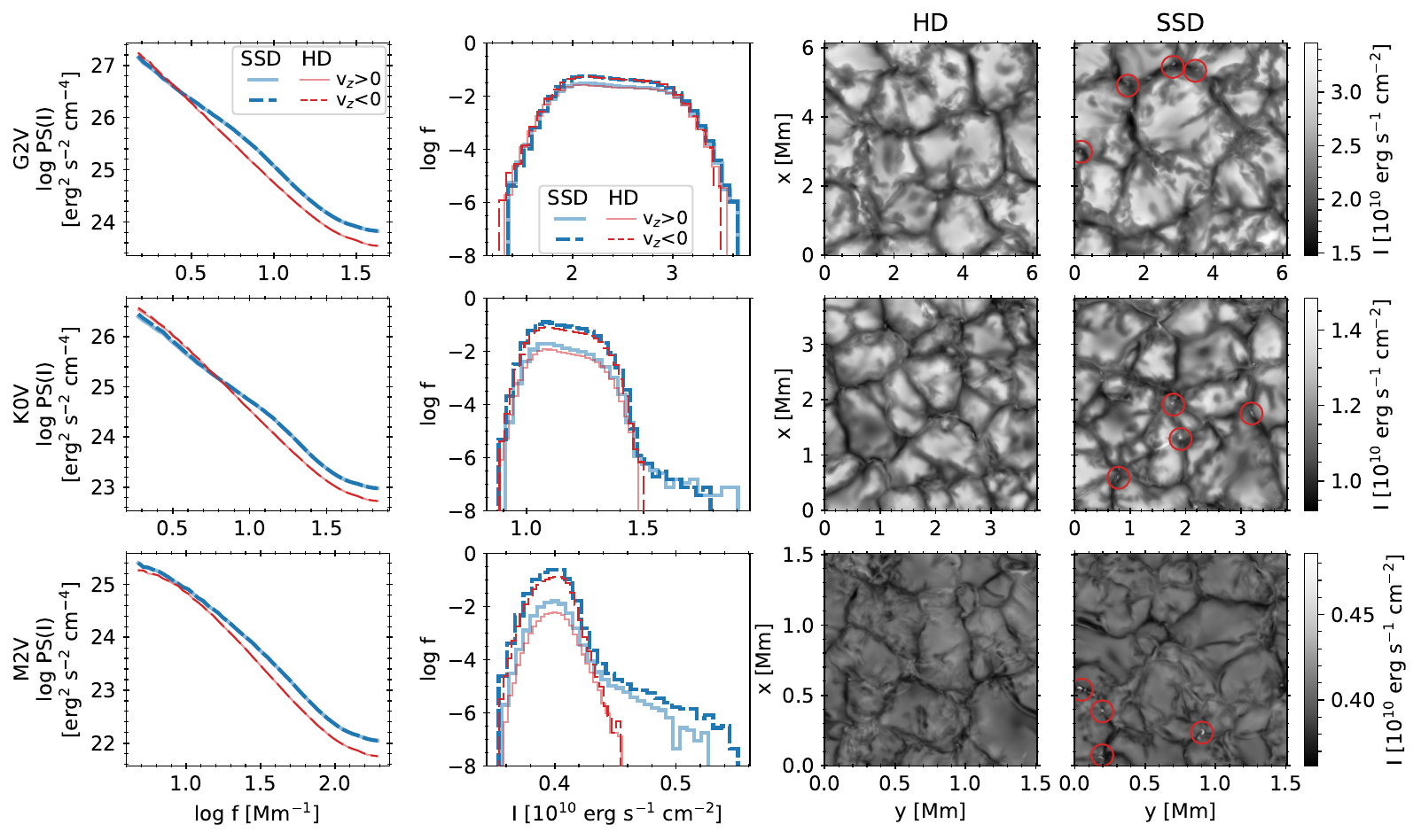}
        \caption{Logarithm of power spectrum, normalized histogram, and colour-maps of the vertical outward bolometric intensity $I$ at the top of the atmosphere for the G2V (top row), K0V (middle), and M2V (bottom) stars. Left column: logarithm of the time averaged radial spatial power spectrum of the intensity versus the logarithm of the radial spatial frequency $f$. Second column: logarithm of the relative frequency $f$ of the normalized histogram of the intensity for the complete time series. In both the left and second columns, dark solid and light dashed lines indicate upflos and downflows, respectively; blue thick and red thin lines indicate SSD and HD runs, respectively. 
        Third and right columns: colour-maps of the intensity of the last snapshot of the HD (third column) and SSD (right) time series. Red circles show the location of magnetic bright points produced by flux concentrations (see Fig. \ref{fig:2D_magnetic_field_ssd}).}
        \label{fig:intensities}
    \end{figure*}

    \begin{figure*}[t!]
        \centering
        \includegraphics[width=14cm]{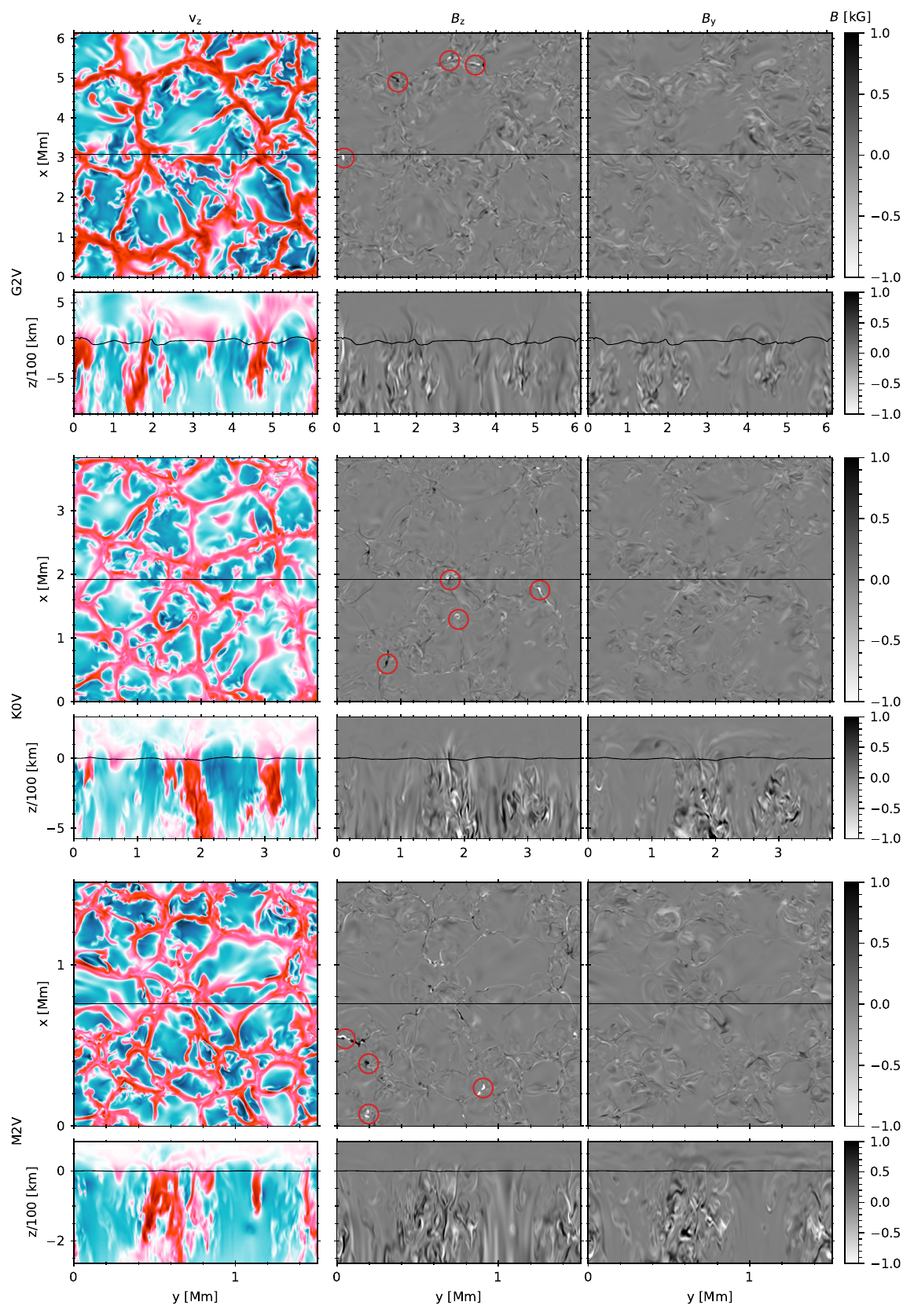}
        \caption{Vertical $B_z$ (middle column) and horizontal $B_y$ component (right column) of the magnetic field of the last snapshot of the SSD time series of the G2V (top pair of rows), K0V (middle), and M2V (bottom) stars. The vertical velocity (left column) is shown again to facilitate the visualization of the magnetic fields in terms of the velocity structure. The values of the magnetic fields shown in the colour-maps are indicated by the corresponding colour-bars (all of them clipped at $|B|=1$ kG). The upper panel of each pair of rows shows the surface at $\tau=1$; the lower panel shows a vertical cut at the middle of the snapshot. The location of the iso-$\tau$ surface and the vertical cut are shown as black solid curves. Red circles show the location of flux concentrations that produce magnetic bright points (see Fig. \ref{fig:intensities}).} 
        \label{fig:2D_magnetic_field_ssd}
    \end{figure*}

    Figures \ref{fig:2D_quantities_ssd_hd_1} and \ref{fig:2D_quantities_ssd_hd_2} (runs with \obc\ opacity setup) are consistent with the description in figure 5 from \citetalias{2024paperII} (runs with Rosseland grey opacity).  
    As in \citetalias{2024paperII}, Figs. \ref{fig:2D_quantities_ssd_hd_1} and \ref{fig:2D_quantities_ssd_hd_2} show how the granulation is qualitatively different for the three stars, with the size of the granules and the smoothness of the granular edges both decreasing gradually from the G2V to the M2V star. The corrugation of the $\tau=1$ surface is also decreasing from the G2V to the M2V star, as well as the temperature contrast, the corrugation of the isotherms, and the strength of the convective velocities. For the three stars, the cooling concentrates below $\tau=1$, while the largest values of the heating appear in the intergranular lanes, although the heating that finally contributes significantly to the stratification concentrates above $\tau=1$.
    
    Although in most aspects Figures \ref{fig:2D_quantities_ssd_hd_1} and \ref{fig:2D_quantities_ssd_hd_2} are consistent with figure 5 from \citetalias{2024paperII}, there are still some differences.  
    The density at $\tau=1$ from the K0V and M2V star and the temperature at $\tau=1$ from the M2V star are significantly lower in the grey runs (figure 5 from \citetalias{2024paperII}) than in the 4-bins runs (Figs. \ref{fig:2D_quantities_ssd_hd_1} and \ref{fig:2D_quantities_ssd_hd_2}). A subtle difference observed in the M2V star is that simulation runs with \obc\ setup present slightly larger values of the heating above the surface (bottom right panels of Fig. \ref{fig:2D_quantities_ssd_hd_2}) than the simulations run with Rosseland grey opacity (bottom right panel of figure 5 from \citetalias{2024paperII}). 

    A more quantitative description of how similar the granulation between the SSD and HD runs is given in Figure \ref{fig:granule_size_ssd_vs_hd}, where the actual variation of the relative area covered by upflows with height is shown in the top panel for the three stars. 
    In the bottom panel, the difference between the SSD and HD cases is shown. Under the surface, the difference is slightly negative for the three stars and becomes almost zero above the surface. Around $\tau=1$ the difference of the relative area covered by upflows is $\lesssim 0.006$. Above the surface, the uncertainty of the difference increases, but it is still an order of magnitude lower than the area itself.

    More significant differences between the granulation of SSD and HD runs are expected for the outward bolometric intensity $I$ (computed column by column). The logarithm of the spatial power spectrum of $I$ is shown at the left column of Figure \ref{fig:intensities} for upflows and downflows. 
    The spatial power spectrum is obtained as described in Appendix \ref{app:FT}.  
    While the power spectrum of $I$ does not show evident differences between the SSD and HS runs, the impact of the SSD on the intensity is evinced in the histograms of $I$ (second column in Fig. \ref{fig:intensities}) and the magnetic bright points found in the SSD runs (compare the third and right column of Fig. \ref{fig:intensities} and see the locations marked with red circles). The magnetic bright points appear at the location of flux concentrations of vertical magnetic field in the intergranular lanes (compare right column of Fig. \ref{fig:intensities} and middle column of Fig. \ref{fig:2D_magnetic_field_ssd}), 
    where the evacuation of material due to pressure balance deepens the iso-$\tau$ surfaces and reveals hotter and brighter plasma compared to the surroundings. 
    These bright points have been extensively observed and simulated for the Sun (see e.g. \citealp{2004SA_BP} and \citealp{2017muram_brightPoints}) and previous simulations for other stars in \citet{2022bhatia_ssd} and \citet{2024Riva_ssd}. There are noticeable tails in the histogram of $I$ of the SSD runs compared to the HD ones: the intensity in the SSD snapshots reaches larger values than in the HD runs, particularly for the downflows in the K0V and M2V stars and the upflows for the G2V and M2V stars.

    \subsection{Properties of the small-scale magnetic fields}

    \begin{figure*}
        \centering
        \includegraphics[width=16cm]{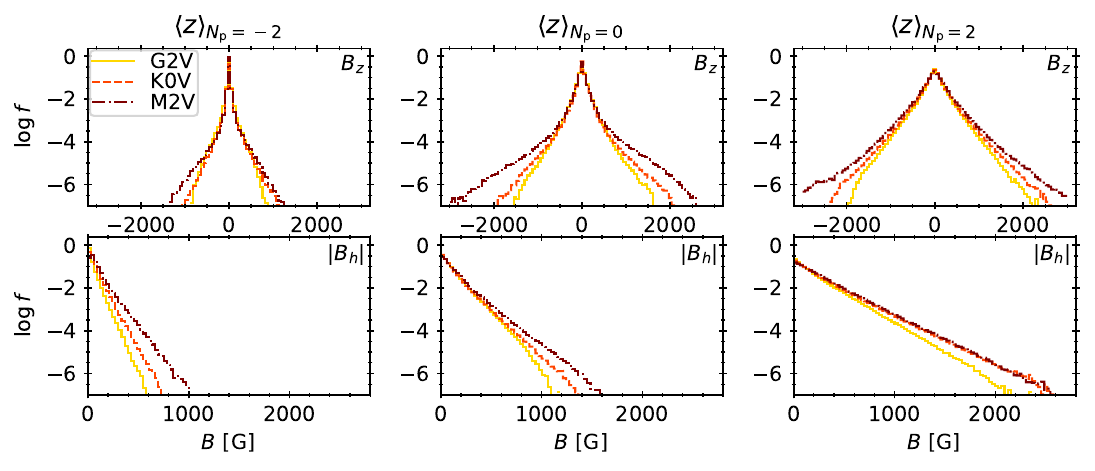}
        \caption{Normalized histograms of the amplitude of the vertical (top row) and horizontal (bottom) component of the magnetic field for the three stars (G2V in yellow solid line, K0V in orange dashed, and M2V in brown dot-dashed). The logarithm of the relative frequency $f$ is shown for the mean geometrical height of $N_\mathrm{p}=-2$ (left column), $N_\mathrm{p}=0$ (center), and $N_\mathrm{p}=2$ (right). }
        \label{fig:histogram_B}
    \end{figure*}

    \begin{figure*}
        \centering
        \includegraphics[width=16cm]{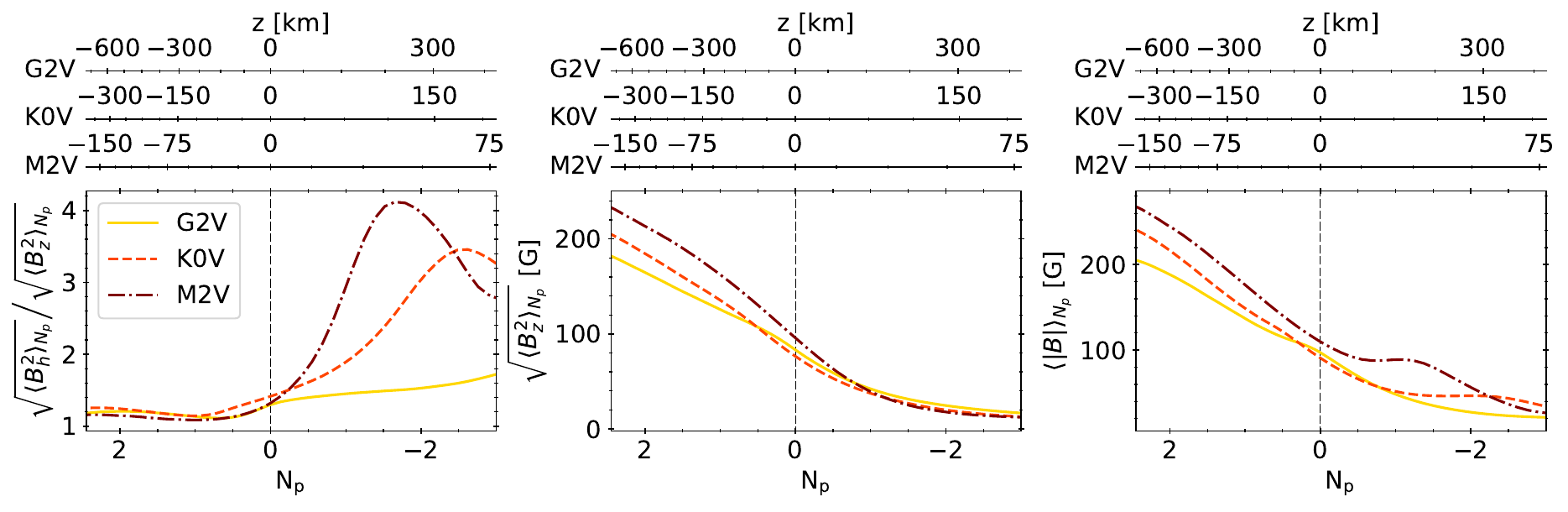}
        \caption{Ratio of RMS horizontal and RMS vertical magnetic field components (left panel), RMS vertical magnetic field component (middle), and modulus of the magnetic field (right) for the three stellar SSD simulations (G2V in yellow solid line, K0V in orange dashed, and M2V in brown dot-dashed) averaged over time and surfaces of constant number of pressure scale heights $N_\mathrm{p}$ (see Eq. \ref{eq:np}). Three top axes on each column: geometrical height for the three stars. }
        \label{fig:1D_magnetic_field_ssd}
    \end{figure*}

    \begin{figure}
        \centering
        \includegraphics[width=8.8cm]{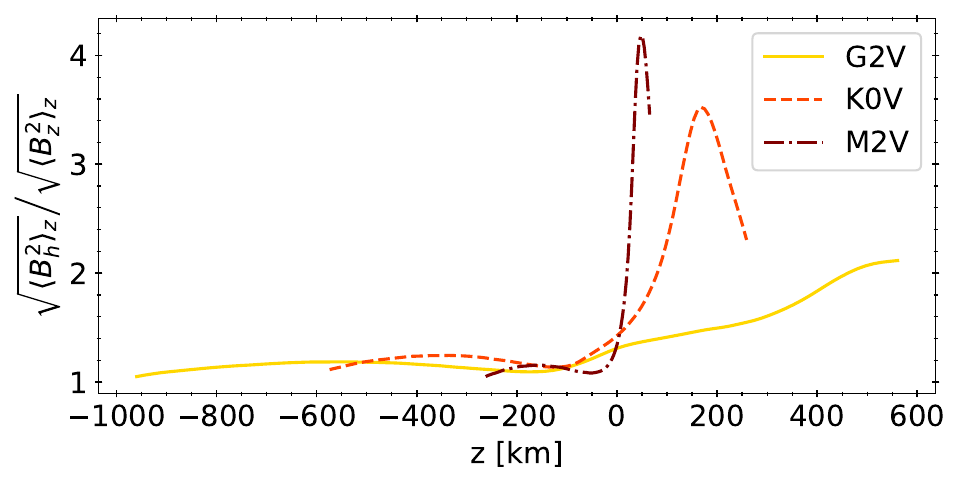}
        \caption{Same as left panel in Fig.\ \ref{fig:1D_magnetic_field_ssd} (ratio of RMS horizontal and RMS vertical magnetic field components), but averaged for surfaces of constant geometrical height. }
        \label{fig:1D_magnetic_field_ZZ_ssd}
    \end{figure}

    The saturated field is shown as coloured-maps in Fig.\ \ref{fig:2D_magnetic_field_ssd}. The figure shows the vertical component and one of the horizontal components of the magnetic field in the $\tau=1$ surface and a vertical cut of the snapshot for the three stars.  
    The vertical velocities are shown to facilitate the comparison with respect to the velocity structure. Consistent with the literature (see e.g. the review by \citealp{2023Rempel_reviewSSD}), the magnetic field is concentrated in the downflowing intergranular lanes and it is stronger below the optical surface. Below the surface, the horizontal fields have the largest values in the edges of the downflows (where the field is more turbulent), and the vertical fields are the strongest within the downflows (see Fig.\ \ref{fig:correlations} in Appendix \ref{app:correlations}). 
    To better understand the distribution of the magnetic fields, the normalized histograms of the amplitude of the vertical and horizontal component of the magnetic field are shown in Fig.\ \ref{fig:histogram_B} for the three stars and the mean geometrical height of $N_\mathrm{p} = -2, 0, 2$. The horizontal component of the magnetic field is computed as $B_h = \sqrt{B_x^2+B_y^2}$. 
    The distribution of vertical magnetic field $B_z$ (top row) is similar for the three stars for $B_z \in [-700, 700]$ G. 
    In the three studied heights, the flux imbalance for the signed $B_z$ is negligible. 
    It has been discussed regarding observations of the Sun how the small dependence on the solar cycle and latitude of the magnetic flux imbalance in the quiet sun is consistent with the SSD action (although this does not discard some small influence from the large-scale magnetic fields, see e.g. \citealp{2011lites_hinode_ssd, 2014lites_hinode_ssd}). 
    The distribution of horizontal magnetic field $B_h$ (bottom row) is similar for the three spectral types for $|B_h| < 700$ G at $N_\mathrm{p} = 0, 2$. For larger $|B_h|$ and at $N_\mathrm{p} = -2$ the distribution of magnetic fields differs between the stars. At $N_\mathrm{p}=-2,0$, values of $|B_h|>700$ G are more frequent in the M2V star, followed by the K0V star, and the G2V star. At $N_\mathrm{p} = 2$, the distribution of $|B_h|$ for the M2V and K0V stars is very close for all $|B_h|$.  
    Following this analysis, the distributions in Fig.\ \ref{fig:histogram_B} show that surface magnetic concentrations are more frequent and have larger values of magnetic fields in the M2V star, followed by the K0V star, and G2V star. 

    Figure\ \ref{fig:1D_magnetic_field_ssd} shows mean properties of the stratification of these small-scale magnetic fields. 
    The left panel shows the mean stratification of the ratio of the RMS horizontal and RMS vertical component of the magnetic field. The ratio is close to unity below the surface. Since an isotropic magnetic field would give a ratio of $\sqrt{2}$, this means the magnetic field is close to isotropic in average, although the vertical component is slightly more dominant than the horizontal one. This result agrees with figure 2 from \citet{2022bhatia_ssd}. Above the surface, the ratio increases significantly for the three stars, peaking at $N_\mathrm{p} \simeq -2.5$ and $N_\mathrm{p} \simeq -2$ for the K0V and M2V star, respectively, implying that the field is mainly horizontal at these heights. This transition from dominant vertical to horizontal component in the magnetic field can be 
    consistent with magnetic flux-tubes or small-scale magnetic loops (e.g. \citealp{2007MJ, 2008Lites, 2015Kitiashvili}, for the Sun). 
    The peak of the ratio of the RMS horizontal and RMS vertical component of the magnetic field does not appear in the case of the G2V star. To understand this behaviour for the G2V star, 
    Fig.\ \ref{fig:1D_magnetic_field_ZZ_ssd} shows the same as in the left panel of Fig.\ \ref{fig:1D_magnetic_field_ssd}, but against the geometrical scale. We can see how the imposition of vertical magnetic field in the top boundary (see Sect. \ref{subsec:convection_mancha}) reduces drastically the ratio at the top of the atmospheres. In the case of the K0V and M2V stars, the peaks of the ratio are located deep down enough so they are not removed by the top boundary condition. In the case of the G2V star, the domain is too shallow and the top boundary makes it very difficult to know the layer where the field becomes horizontal (although it seems that the peak could be located around 500 km above $\tau=1$, as suggested by \citealp{2008steiner} or \citealp{2014rempel_SSD}).

    The middle panel of Fig.\ \ref{fig:1D_magnetic_field_ssd} shows the RMS vertical component of the magnetic field for the three stars. The vertical magnetic field in the three stars decreases with height, ranging from $\simeq 200$ G at the bottom of the domain down to $\simeq 20$ G at the top, with similar stratifications for the three spectral types. Likewise, the modulus of the magnetic field (right panel of Fig.\ \ref{fig:1D_magnetic_field_ssd}) also decreases with height, again displaying similar variations for the three stars. 
    The fact that the heights where the horizontal component of the magnetic field dominates most over the vertical component change between the spectral types (left panel of Fig.\ \ref{fig:1D_magnetic_field_ssd}) can be also observed in the modulus of the magnetic field (right panel of Fig.\ \ref{fig:1D_magnetic_field_ssd}). The modulus of the magnetic field shows humps with locally more intense magnetic field at $N_\mathrm{p} \simeq -1.5$ and $N_\mathrm{p} \simeq -2.5$ for the M2V and K0V, respectively, while no peaks are observed in the RMS vertical magnetic field. 

    Below the surface, the amplitude of the magnetic field (right panel of Fig.\ \ref{fig:1D_magnetic_field_ssd}) is different between the three stars, with up to 40 G and 60 G of difference between the field from the K0V and M2V, respectively, compared to the field of the G2V star. Close to the stellar surface, these differences are reduced and the three stars have fields of the order of 100 G (with the differences lower than 20 G, for any pair of stars). This similarity at $\tau=1$ of the saturation values of the magnetic field amplified by the action of SSD in stars with different spectral types is consistent with the results shown in \citet{2022bhatia_ssd} and \citet{2024Riva_ssd}. There are differences of the amplitude of the magnetic field in sub-surface layers in the results presented in \citet{2022bhatia_ssd} too (the reference treats the same spectral types as in the present paper), although the values are not the same as ours. While in our case the M2V star has the largest value of $|B|$ at $N_\mathrm{p}=2$, followed by the K0V star, followed by the G2V star, at the same height in figure 2 from \citet{2022bhatia_ssd} the K0V and M2V have almost identical $|B|$, larger than the $|B|$ from the G2V star.

    \subsection{Effect on the power spectrum of temperature and velocities}

    \begin{figure*}
        \centering
        \includegraphics[width=16cm]{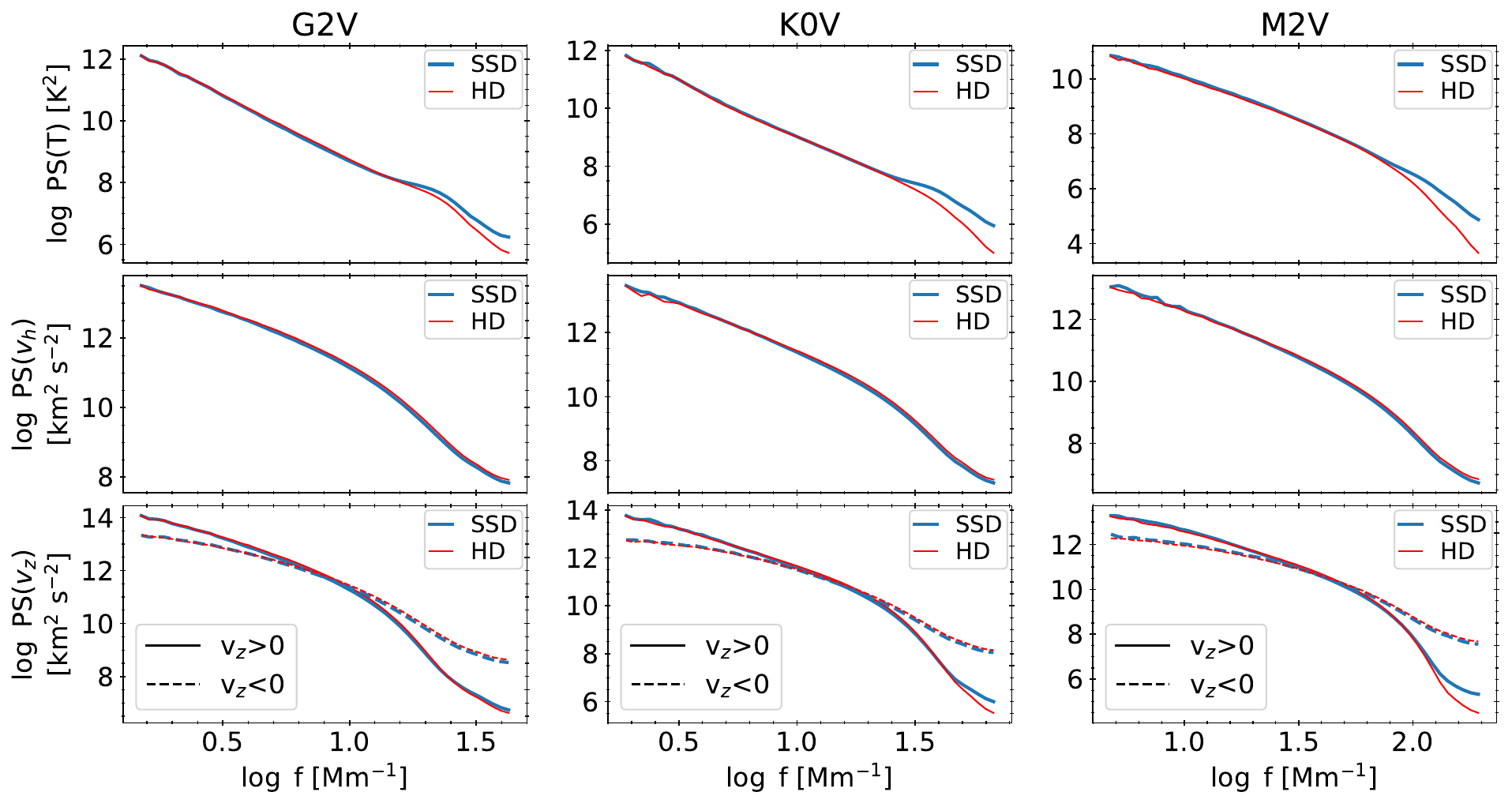}
        \caption{Logarithm of the time averaged radial spatial power spectrum at $\tau=1$ surface of the temperature (top row), horizontal velocity (middle), and vertical velocity (bottom) of upflows (solid line) and downflows (dashed) versus the logarithm of the radial spatial frequency $f$ for the three stars (different columns) and the SSD (blue thick lines) and HD (red thin) runs.}
        \label{fig:power_spectrum}
    \end{figure*}

    Differences between the HD and SSD case may appear in the smaller scales owing to the turbulent magnetic field. 
    Figure \ref{fig:power_spectrum} shows the 
    time averaged radial spatial power spectrum at $\tau=1$ surface of the temperature, amplitude of the horizontal velocity, and amplitude of the vertical velocity of upflows and downflows for the three stars and the SSD and HD runs. 
    The power spectrum is obtained as explained for Fig. \ref{fig:intensities}. 
    The distribution of temperatures on scales with spatial frequencies $f>20 \, \mathrm{Mm}^{-1}$ (G2V; scales of around 50 km), $f>30 \, \mathrm{Mm}^{-1}$ (K0V; 33 km), $f>100 \, \mathrm{Mm}^{-1}$ (M2V; 10 km) is different between the HD and SSD runs, but remains equal for the rest of frequencies. 
    The vertical velocity of upflows for $f>30 \, \mathrm{Mm}^{-1}$ (K0V) and $f>100 \, \mathrm{Mm}^{-1}$ (M2V) shows also slightly different power spectra. In the presence of the small-scale magnetic field, both the temperature and vertical velocity of upflows may either have larger values than the HD case for small structures or more abrupt spatial changes. Further studies with simulations with higher spatial resolution are needed to fully characterize and understand these differences, since the mentioned scales correspond to approximately three grid points in the stellar domains. As expected, downflows are concentrated in smaller structures (intergranular lanes) than upflows (granules), as indicated by the larger (smaller) power in high (low) frequencies. Surprisingly, no significant differences are observed in the power spectrum of the vertical velocity of downflows. The power spectrum of the horizontal velocity is also almost unaffected.

    \subsection{Effect on mean stratification of temperature}

    \begin{figure}[t!]
        \centering
        \includegraphics[width=8.8cm]{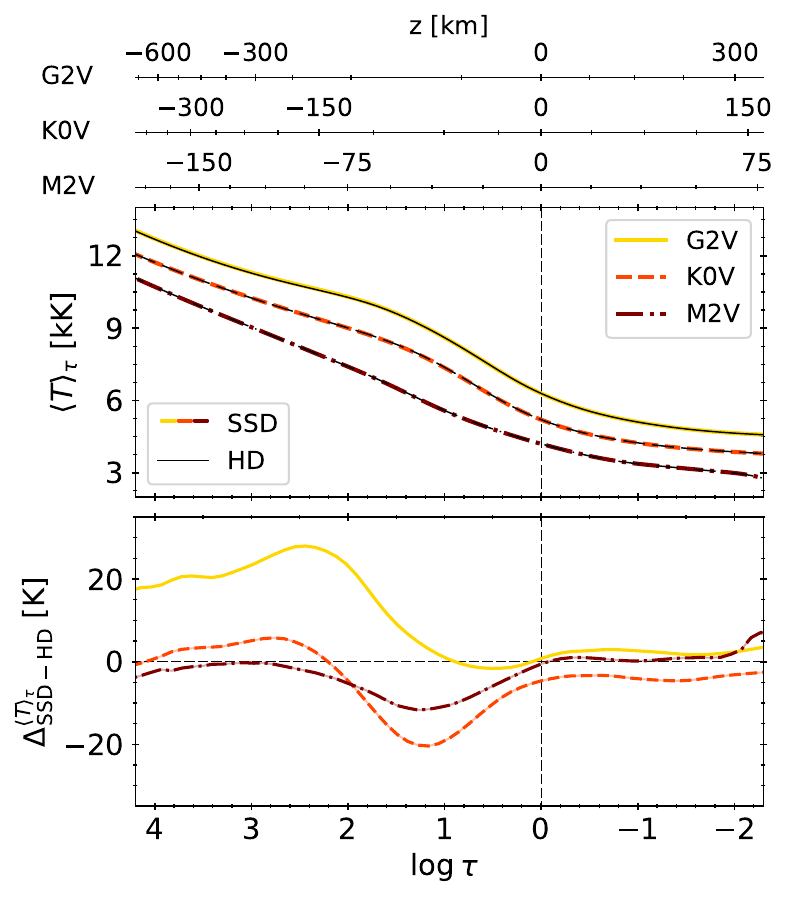}
        \caption{Temperature averaged over time and Rosseland iso-$\tau$ surfaces for the three stars. 
        Similarly to Fig.\ \ref{fig:granule_size_ssd_vs_hd}, the temperature  is shown in the top row (with the values from the SSD and HD simulations in thick coloured and thin black lines, respectively) and the differences between the values from the SSD and HD simulations (Eq.\ \ref{eq:diff_delta}) are shown in the bottom row, with the errors as shaded areas. 
        Three top axes: geometrical height for the three stars.}
        \label{fig:temp_ssd_vs_hd}
    \end{figure}

    \begin{figure}[t!]
        \centering
        \includegraphics[width=8.8cm]{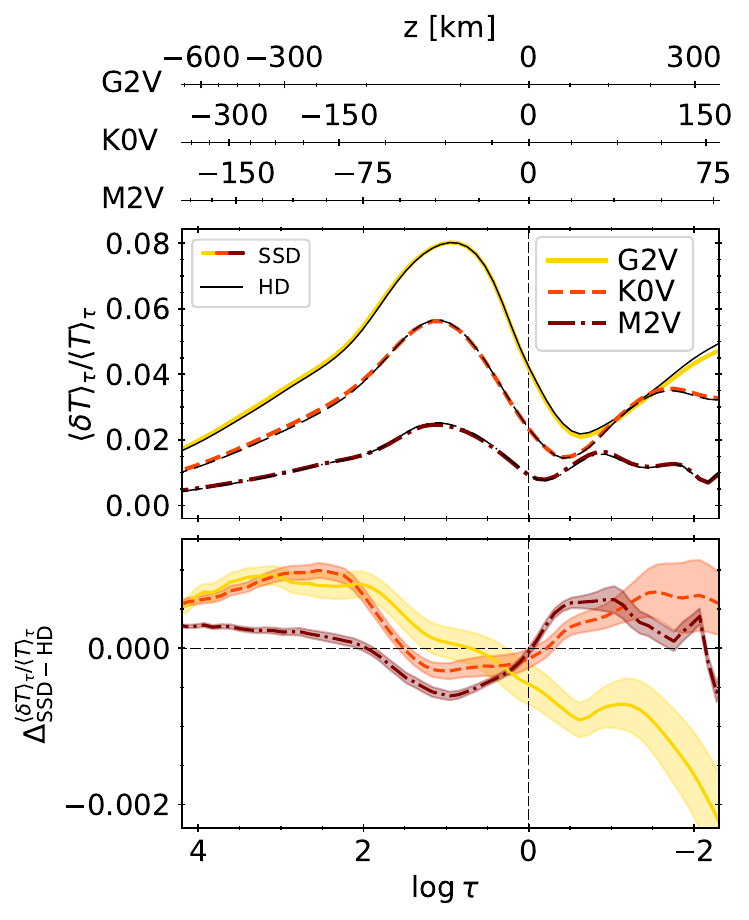}
        \caption{Temperature contrast averaged over Rosseland iso-$\tau$ surfaces and time for the three spectral types. The contrast is computed as the standard deviation of the temperature divided by its mean. 
        Similarly to Fig.\ \ref{fig:granule_size_ssd_vs_hd}, the temperature contrast is shown in the top row (with the values from the SSD and HD simulations in thick coloured and thin black lines, respectively) and the differences between the values from the SSD and HD simulations (Eq.\ \ref{eq:diff_delta}) are shown in the bottom row, with the errors as shaded areas. 
        Three top axes: geometrical height for the three stars. 
        }
        \label{fig:temp_contr_ssd_vs_hd}
    \end{figure}

    \begin{figure}[t!]
        \centering
        \includegraphics[width=8.8cm]{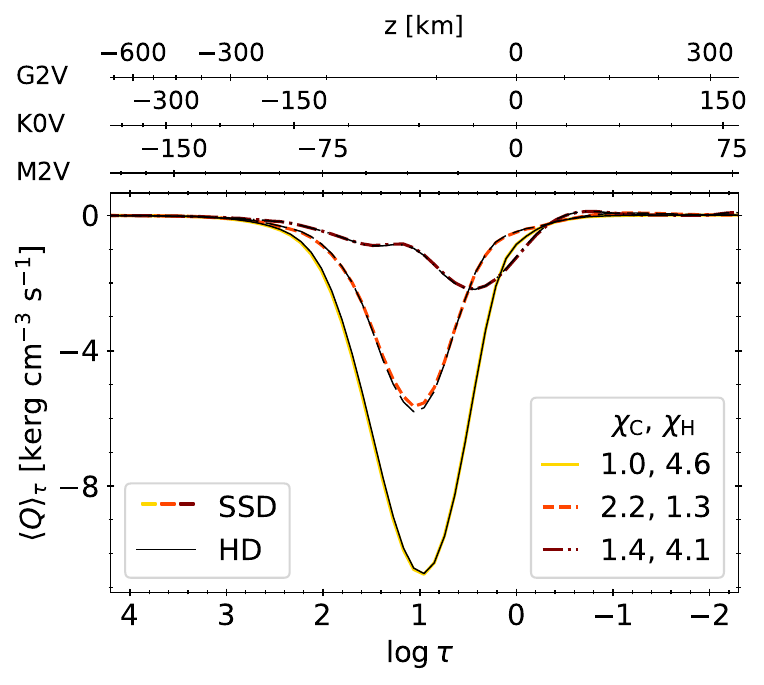}
        \caption{Energy exchange rate $Q$ averaged over time and Rosseland iso-$\tau$ surfaces for the three stars. 
        Similarly to Fig.\ \ref{fig:granule_size_ssd_vs_hd}, the values from the SSD and HD simulations are shown in thick coloured and thin black lines, respectively. 
        The deviations $\chi$ (Eqs.\ \ref{eq:error_c} and \ref{eq:error_h}, with $Q^{(1)}=Q^{\mathrm{SSD}}$ and $Q^{(2)}=Q^{\mathrm{HD}}$) are displayed in the legend. 
        Three top axes: geometrical height for the three stars.
        }
        \label{fig:Q_ssd_vs_hd}
    \end{figure}

    Figure \ref{fig:temp_ssd_vs_hd} compares the mean temperature stratification between the SSD and HD runs for the three stars. The results are very close for the SSD and HD runs, with differences lower than 30, 25, and 15 K in the G2V, K0V and M2V stars, respectively. These differences in temperature are smaller than the differences between the runs with 18 opacity bins and those with four bins or Rosseland grey opacity (see section 4.2 in \citetalias{2024paperII}). The simulations run with grey opacity have more than $150$ K of difference compared to the ones run with the 18-bins opacity. The simulations run with 4-bins opacity reach at some heights up to 100, 150, and 50 K of difference compared to the ones run with the 18-bins opacity for the G2V, K0V, and M2V star, respectively. 

    Figure \ref{fig:temp_contr_ssd_vs_hd} shows the comparison of the mean stratification of the temperature contrast between the SSD and HD runs. For the three stars, in subsurface layers the temperature contrast computed from the SSD simulations is slightly larger than that computed from the HD simulations. Going from the bottom upward, the absolute value of the differences is reduced close to the surface, while it increases again above it. In these layers, the temperature contrast is larger in the HD runs than in the SSD ones for the G2V, while the opposite is true for the other two stellar types. In all stars, the difference is  an order of magnitude lower than the contrast itself. 

    \subsection{Effect on mean stratification of $Q$ rate}

    Figure \ref{fig:Q_ssd_vs_hd} shows the comparison of the mean stratification of the radiative energy exchange rate between the SSD and HD runs for the three stars. The deviations $\chi$ used in \citetalias{2023paperI} and \citetalias{2024paperII}
    \begin{equation} \label{eq:error_c}
        \mathrm{\chi_{\mathrm{C}}} = \frac{ A \left( \left | Q^{(1)}  -  Q^{(2)} \right |  \right)_{\mathrm{C}} }{A \left( \left | Q^{(1)}  \right | \right)_{\mathrm{C}}},
    \end{equation} 
    \begin{equation} \label{eq:error_h}
        \mathrm{\chi_{\mathrm{H}}} = \frac{ A \left( \left | Q^{(1)}  -  Q^{(2)} \right |  \right)_{\mathrm{H}} }{A \left( \left | Q^{(1)}  \right | \right)_{\mathrm{H}}},
    \end{equation}
    are indicated in the figure, 
    with $Q^{(1)}=Q^{\mathrm{SSD}}$ and $Q^{(2)}=Q^{\mathrm{HD}}$. As explained in our previous works, these deviation measures are more convenient than the relative error because they ensure that we do not overestimate the difference of the Q rates in the regions of the atmosphere where the values are close to zero, and avoid the problems that relative error has when Q changes sign close to the surface. The areas are defined as 
    \begin{equation}\label{eq:area_c}
        A \left( f(z) \right)_C = \int_{z_{\mathrm{b}}}^{z_{\mathrm{C \rightarrow H}}} f(z) dz,
    \end{equation}
    \begin{equation}\label{eq:area_h}
        A \left( f(z) \right)_H = \int_{z_{\mathrm{C \rightarrow H}}}^{z_{\mathrm{t}}} f(z) dz.
    \end{equation}
    The height at $z_{\mathrm{C \rightarrow H}}$ is the height where $Q$ changes sign for the first time above the cooling component (see figure 6 in \citetalias{2023paperI}); $z_{\mathrm{b}}$ and $z_{\mathrm{t}}$ are, respectively, the highest point in the atmosphere under the surface ($z<0$) and the lowest point over the surface ($z>0$) where $\left| Q \right|< 2 \times 10^{-4} \left| \min( Q ) \right|$. 

    \begin{table}[b!]
    \caption{Comparison of deviation measures for $Q$.}
    \begin{tabular}{c|cc|cc|cc}
    \hline
                      & \multicolumn{2}{c|}{G2V}              & \multicolumn{2}{c|}{K0V}              & \multicolumn{2}{c}{M2V}               \\ \hline
                      & $\chi_\mathrm{C}$ & $\chi_\mathrm{H}$ & $\chi_\mathrm{C}$ & $\chi_\mathrm{H}$ & $\chi_\mathrm{C}$ & $\chi_\mathrm{H}$ \\ \hline
    \grey\ vs \obd\   & 2.3               & 82.6              & 8                 & 16.6              & 20.8              & 89.1              \\ \hline
    \obc\ vs \obd\ & 1.2               & 22.8              & 1.6               & 12.5              & 3.5               & 38.8              \\ \hline
    SSD vs HD         & 1.0               & 4.6               & 2.2               & 1.3               & 1.4               & 4.1               \\ \hline
    \end{tabular}
    \label{tab:comparison_Qs_ssd}
    \tablefoot{Deviation measures for $Q$ of the mean stratifications computed with the \grey\ (top row) and \obc\ (middle) runs with respect to the \obd\ run from section 4.2 in \citetalias{2024paperII} and the HD runs with respect to the SSD runs (bottom row) from the present section.}
    \end{table}

    Table\ \ref{tab:comparison_Qs_ssd} compares these deviations against the ones produced by the different opacity approaches (compare also Fig.\ \ref{fig:Q_ssd_vs_hd} and top row of figure 19 in \citetalias{2024paperII}). 
    The deviations in $Q$ between the SSD and HD runs are in most cases significantly lower than those corresponding to the \obc\ and \grey\ runs in comparison to the \obd\ run from section 4.2 in \citetalias{2024paperII}. Both the temperature and $Q$ stratifications are more affected by the treatment of the opacity than by the presence of the magnetic fields amplified by the action of SSD.

    \subsection{Effect on mean stratification of velocities}

    \begin{figure*}[t!]
        \centering
        \includegraphics[width=16cm]{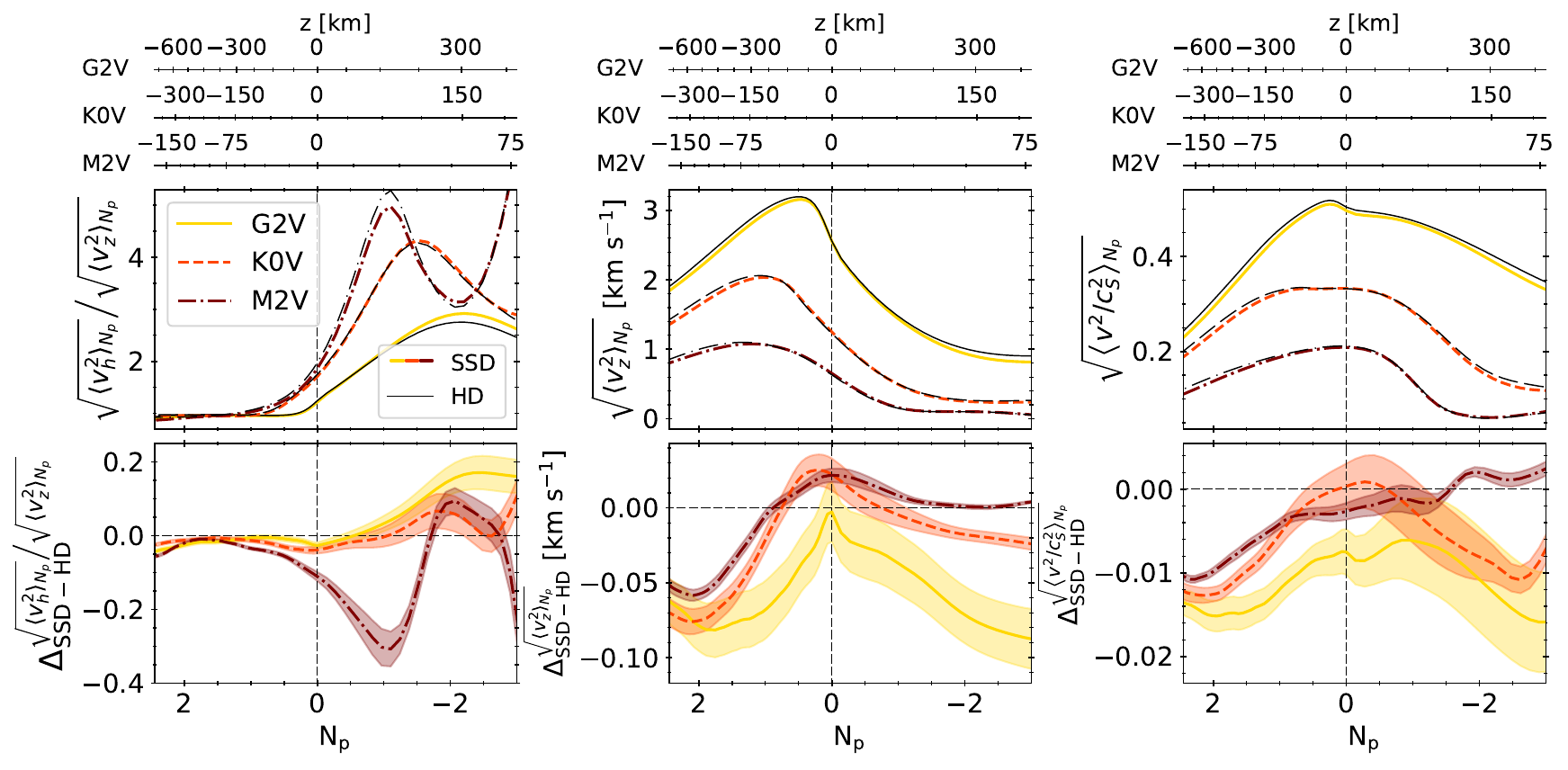}
        \caption{Ratio of RMS horizontal and RMS vertical velocities (left column), RMS vertical velocity (middle), and Mach number (right) averaged over time and surfaces of constant number of pressure scale heights $N_\mathrm{p}$ (see Eq. \ref{eq:np}), versus $N_\mathrm{p}$. Similarly to Fig.\ \ref{fig:granule_size_ssd_vs_hd}, the quantities are shown in the top row (with the values from the SSD and HD simulations in thick coloured and thin black lines, respectively) and the differences between the values from the SSD and HD simulations (Eq.\ \ref{eq:diff_delta}) are shown in the bottom row, with the errors as shaded areas. 
        Three top axes on each column: geometrical height for the three stars. }
        \label{fig:vel_stat_ssd_vs_hd_1}
    \end{figure*}

    Since the spatial power spectra for the velocities show that scales below certain size are affected by the small-scale magnetic field, 
    the mean velocity field may show small changes. Figure\ \ref{fig:vel_stat_ssd_vs_hd_1} (top row) shows mean properties of the velocity field of the three stars as a function of the number of pressure scale heights for the SSD (coloured lines) and HD (black thin lines) cases. The differences between the SSD and HD cases (Eq.\ \ref{eq:diff_delta}) for each quantity are shown in the bottom row. The ratio of RMS horizontal and RMS vertical velocities (top left panel) displays similar trends for the SSD and HD simulations for the three stars. In both types of simulations, the ratio is close to unity for $N_{\mathrm{p}}>1$. The horizontal component of the velocity becomes larger than the vertical for $N_\mathrm{p}<1$ and the ratio peaks at $N_\mathrm{p} \simeq -2$, $N_\mathrm{p} \simeq -1.5$, and $N_\mathrm{p} \simeq -1$ for the G2V, K0V, and M2V stars, respectively. In the case of the G2V star, the peak ratio is larger in the SSD case than in the HD case, while it is the other way around in the M2V star. For the other heights in the atmosphere of these two stars, and all heights in the K0V star, the differences are very small. Actually, the effect of using 4-bins opacity instead of grey produces more evident changes in the position $N_\mathrm{p}$ of the peaks of the K0V and M2V star. The positions from the simulations with four bins ($N_\mathrm{p} \simeq -1.5$ and $N_\mathrm{p} \simeq -1$, respectively; see Fig.\ \ref{fig:vel_stat_ssd_vs_hd_1}) appear in different heights than the grey case ($N_\mathrm{p} \simeq -1.3$ for both stars; see figure 1 in \citetalias{2024paperII}). The non-grey opacity also affects the amplitude of the peak of the M2V star, which is larger in the simulations with four bins ($\simeq 5$; see Fig.\ \ref{fig:vel_stat_ssd_vs_hd_1}) than from the grey simulations ($\simeq 4$; see figure 1 in \citetalias{2024paperII}).

    According to the  middle panels of Fig.\ \ref{fig:vel_stat_ssd_vs_hd_1}, in sub-surface layers the RMS vertical velocity for the HD runs is slightly larger than in the SSD run, which agrees with a situation in which kinetic energy has been transformed into magnetic. 
    Close to the optical surface, the difference between the SSD and HD runs becomes minimum for the three stars. Above the optical surface, the stars show again a larger RMS vertical velocity for the HD runs than for the SSD runs.

    The right panels of Fig.\ \ref{fig:vel_stat_ssd_vs_hd_1} show the Mach number of the three stars for the SSD and HD cases and the differences between the SSD and HD cases. Similarly to the RMS vertical velocity, the differences between the SSD and HD runs for the G2V and K0V stars are minimal close to the surface, becoming larger deeper down and higher up the surface. In the M2V star, the differences are maximal in subsurface layers, and become close to zero around the surface and above. For the three stars, in sub-surface layers the mean Mach number is larger in the HD case than in the SSD case. 
    From the negligible differences of the temperature between the SSD and HD runs (Fig. \ref{fig:temp_ssd_vs_hd}), the mean sound speed (if approximated by the expression for ideal gas) is expected to be virtually the same in the SSD and HD runs.  
    This implies that the lower Mach number in sub-surface layers in the SSD compared to the HD runs is likely due to the reduction in kinetic energy because of the action of the SSD (see figures for the energy stratification at the end of the present section). Although the mean stratification of the Mach number is always below one for the three stars, the G2V and K0V stars still have sporadic supersonic flows in their domain. For the whole domains of both stars, the number of supersonic cells is $1.2$ (G2V) and $5.9$ (K0V) times larger in the HD than in the SSD runs.

    \subsection{Energy transformation}

    \begin{figure*}[t!]
        \centering
        \includegraphics[width=16cm]{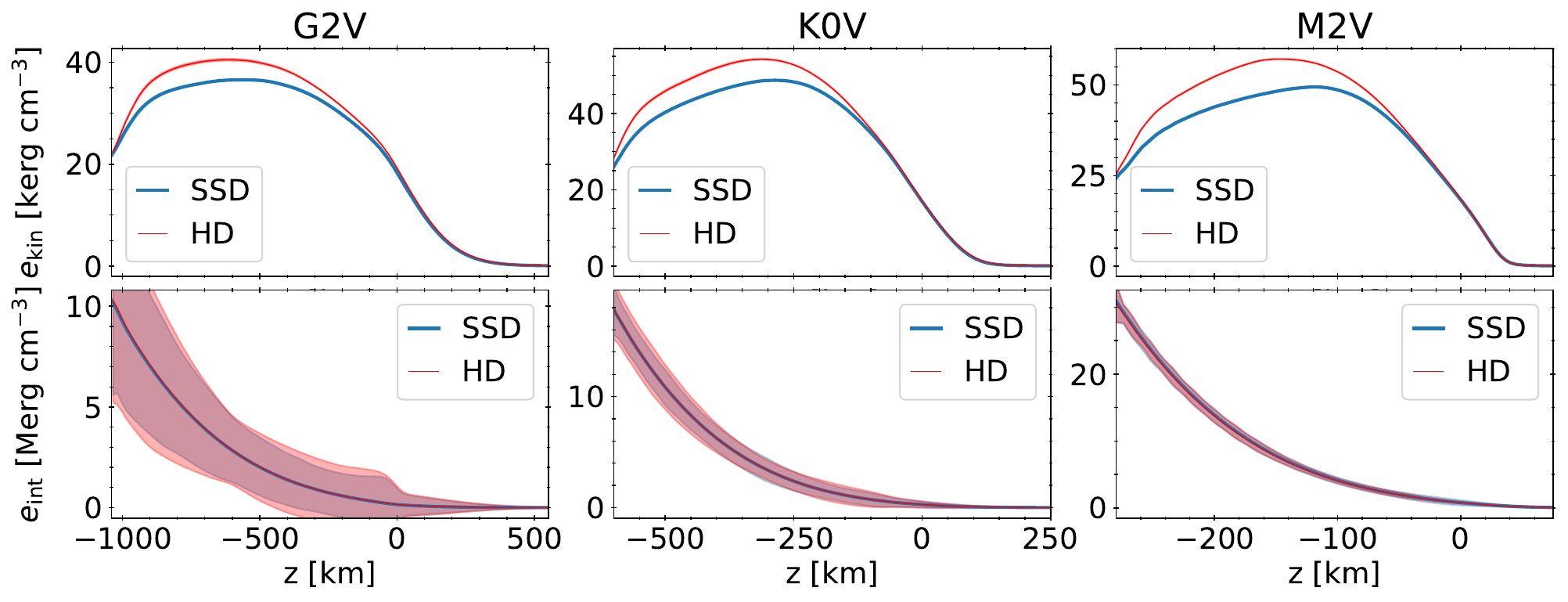}
        \caption{Kinetic (top row) and magnetic (bottom) energy over surfaces of constant geometrical height and time for the three stars (each column). The energy from the SSD (HD) simulations is shown in blue (red) line. Shaded areas show $\sigma/\sqrt{N}$ for the kinetic and internal energy.}
        \label{fig:energies_ssd_vs_hd_strat}
    \end{figure*}

    \begin{figure*}[t!]
        \centering
        \includegraphics[width=16cm]{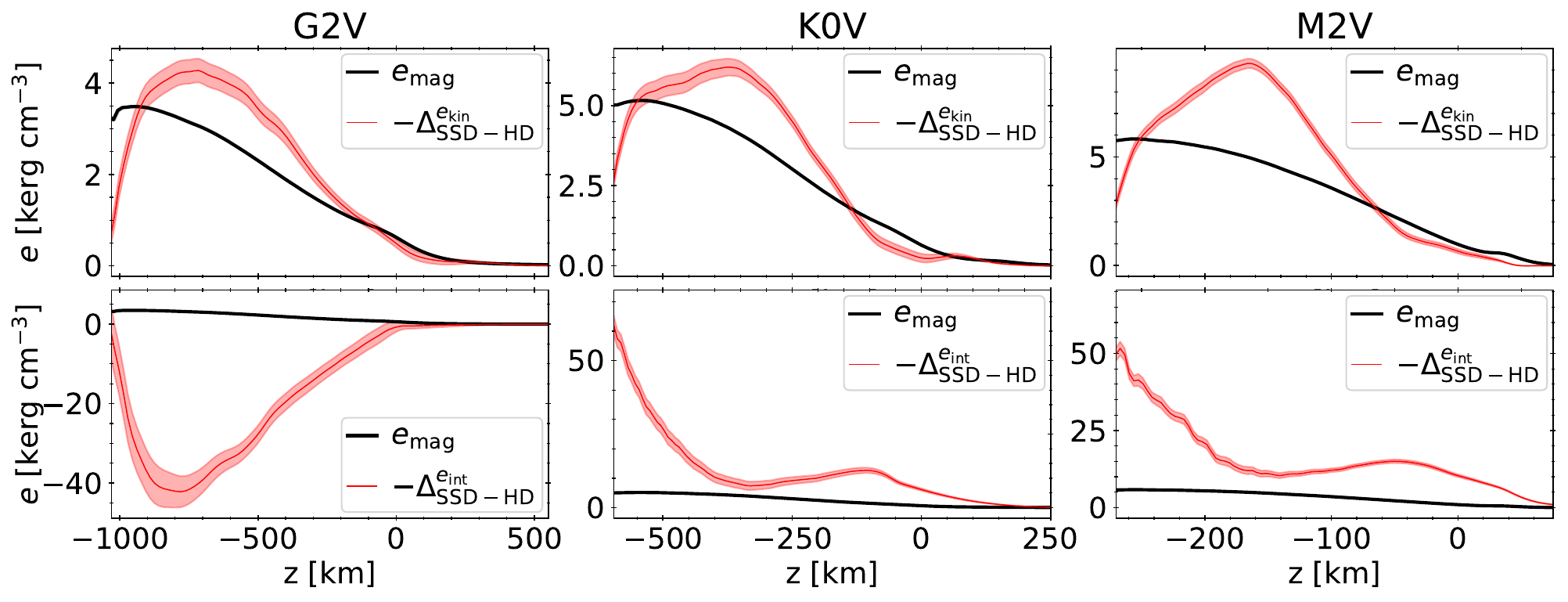}
        \caption{Magnetic energy averaged over surfaces of constant geometrical height and time from the SSD runs (black lines), compared to the difference in kinetic (top row) and internal (bottom) energy (Eq. \ref{eq:diff_delta} multiplied by -1; red lines) between the HD and SSD runs. Shaded areas show $\sigma/\sqrt{N}$ for the magnetic energy and the error (Eq. \ref{eq:error_differences}) for the difference of energy between the SSD and HD cases.}
        \label{fig:energies_ssd_vs_hd_comparison}
    \end{figure*}

    \begin{figure*}
        \centering
        \includegraphics[width=16cm]{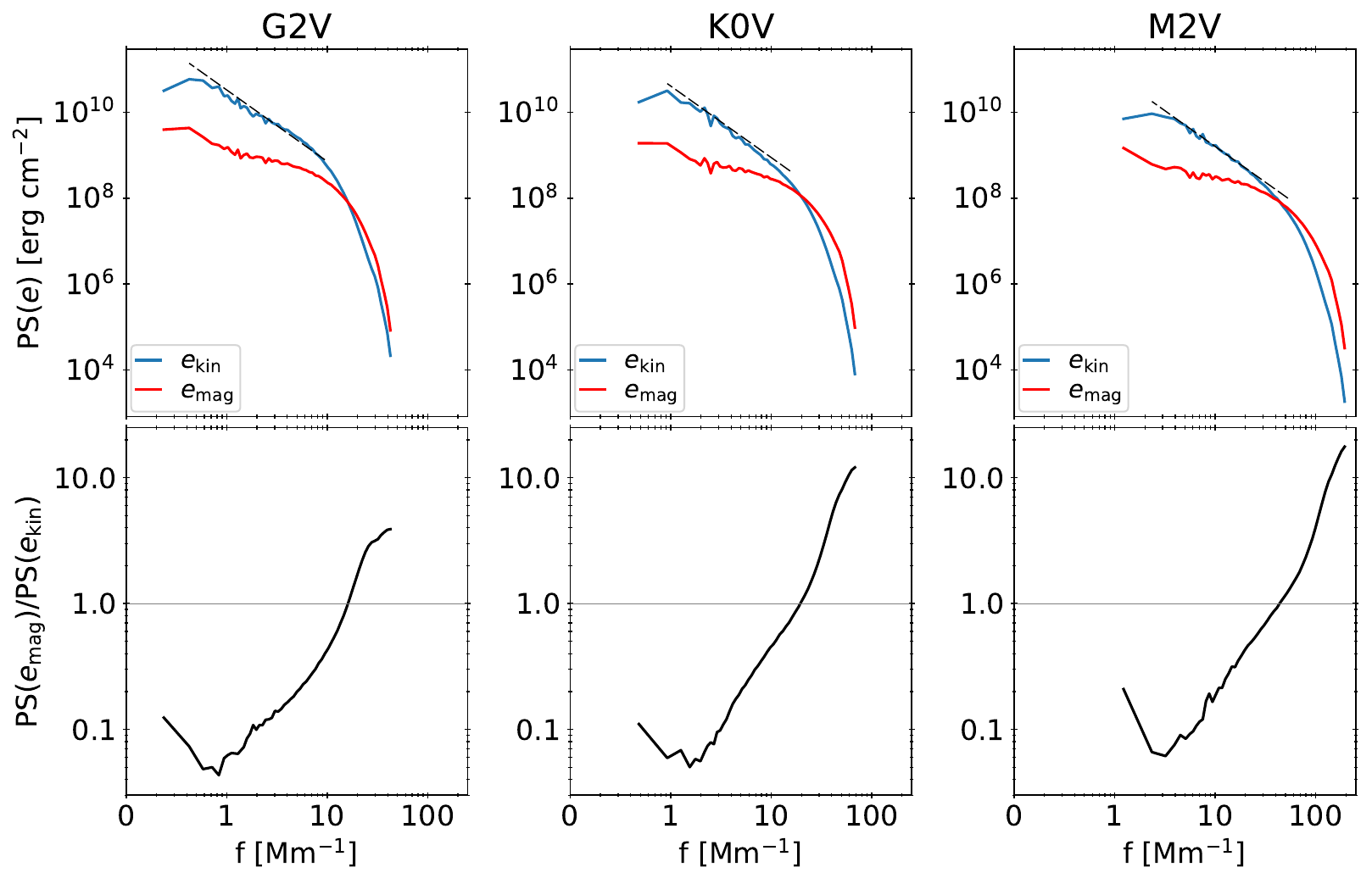}
        \caption{Time averaged radial spatial power spectra (top row) of the kinetic (blue lines) and magnetic (red) energies at $\tau=1$ for the SSD runs of the three stars (different columns). In dashed lines we show the curve $a\times f^{-5/3}$ (Kolmogorov power law, where $a$ is a constant), where $f$ is the radial spatial frequency. The ratio of the spatial power spectra of the magnetic energy and the spatial power spectra of the kinetic energy is shown in the bottom row.}
        \label{fig:energy_power_spectra}
    \end{figure*}

    The energy distribution is expected to change in the SSD runs with respect to the HD runs, due to the transformation of kinetic energy into magnetic energy through the SSD action. 
    Figure \ref{fig:energies_ssd_vs_hd_strat} shows the mean stratification of the kinetic and internal energy for the three stars and the SSD (blue lines) and HD (red) runs. 
    Figure \ref{fig:energies_ssd_vs_hd_comparison} compares the mean stratification of the magnetic energy in the SSD runs (black lines) and the variations of kinetic and internal energy (red) of the SSD runs with respect to the HD runs for the three stars. 
    We have plotted the reverse-signed variation of $\Delta$ (i.e., HD-SSD), rather than the SSD-HD difference used in the rest of the paper (Eq.\ \ref{eq:diff_delta}). This way, the magnitude is essentially positive and facilitates the comparison with the magnetic energy. 
    The three stars show a decrease of kinetic energy in the SSD runs with respect to the HD runs below the surface. 
    Above the surface, the difference in kinetic energy between the SSD and HD runs becomes close to zero. The same happens to the change in the internal energy, which becomes small above the surface. 
    Below the surface, the G2V has more internal energy in the SSD run than the HD run, while the K0V and M2V stars have less internal energy in the SSD run than in the HD run. 
    
    The difference of kinetic energy between the HD and SSD runs (Fig. \ref{fig:energies_ssd_vs_hd_strat}) as a function of height is an order of magnitude smaller than the kinetic energy itself (Fig. \ref{fig:energies_ssd_vs_hd_comparison}), while the difference of internal energy between the HD and SSD runs is two orders of magnitude smaller than the internal energy itself. Thus, while the energy difference for the internal energy is negligible, a substantial amount of kinetic energy is lost when comparing the HD and SSD case. 
    
    In summary, the difference in kinetic energy has the same order of magnitude and similar variation with height (decreasing for the upper $2/3$ of the domain) as the magnetic energy. This means that there is a correspondence between the stratifications of the magnetic energy and the difference in kinetic energy. On the contrary, the difference in internal energy is significantly larger than the magnetic energy and shows different variation with height: either increasing (G2V), or non-monotonic (K0V and M2V).

    Finally, Fig.\ \ref{fig:energy_power_spectra} shows the radial spatial power spectra of the kinetic and magnetic energies at $\tau=1$ averaged in time for the SSD run of the three stars. The spectra are computed as detailed in Appendix \ref{app:FT}.
    
    The inertial range of the turbulence is the range of spatial frequencies between the maximum scale of injection of energy from larger scales to smaller scales and the maximum scale of dissipation of energy due to viscosity. The power spectra in Fig.\ \ref{fig:energy_power_spectra} shows a dependence of the inertial range and the maximum scale for super-equipartition between kinetic and magnetic energy on the spectral type. The inertial range is at $f \in [0.5, 10]$ Mm$^{-1}$ (for scales between 100 km and 2 Mm), $f \in [1, 20]$ Mm$^{-1}$ (between 50 km and 1 Mm), and $f \in [3, 50]$ Mm$^{-1}$ (between 20 km and 300 km) for the G2V, K0V, and M2V, respectively. The super-equipartition happens for scales lower than 60 km for the G2V star, 50 km for the K0V star, and 20 km for the M2V star. 
    \citet{2014rempel_SSD} shows for the Sun that the maximum scale for super-equipartition between kinetic and magnetic energy depends on the resolution: in the middle panel of their figure 3, this scale has the values of approximately 120, 109, 104, and 90 km (equivalent to around 8, 14, 26, and 46 grid points) for the corresponding cell-sizes of 16, 8, 4, and 2 km. The minimum scale of dissipation of energy has similar values to those of the equipartition scale. 
    In principle these scales are expected to vary with resolution for other stars, but our study is limited to one resolution. Nevertheless, since our three stars have equipartition scales that include a similar number of grid points (between five and seven), our evidence suggests that the equipartition scale and the minimum scale of the inertial range depend indeed on the spectral type.

\section{Conclusions} \label{sec:conclusions}

    In the present work, we continue the simulations with the \texttt{MANCHA} code presented in \citetalias{2024paperII} for G2V, K0V, and M2V stellar atmospheres with solar metallicity, after adding the Biermann battery term in the induction equation. The resulting simulations contain small-scale magnetic fields and are analysed putting our results in context with previous works and using our HD simulations as a reference to measure the impact of the SSD in different stars.
    
    The growth of the magnetic field with time (Fig. \ref{fig:saturation_vs_time}) shows the phases described previously in the literature (e.g. \citealp{2007voegler_Schussler_SSD} and \citealp{2017khomenko}): a first phase of quasi-linear growth of the field, followed by the SSD amplification until saturation.

    The three simulated stars reach similar saturation values of the magnetic field strength of around 100 G at $\tau=1$, agreeing with the values found in \citet{2022bhatia_ssd} and \citet{2024Riva_ssd}. 
    For the three stars, the mean stratification of the ratio of the RMS horizontal and RMS vertical magnetic field show that the magnetic field is close to isotropic but still dominated by its vertical component below the surface and gets progressively more dominated by its horizontal component while the height grows. The ratio peaks above the surface, where the field is mainly horizontal. This description is 
    similar to that shown in \citet{2014rempel_SSD} for the case of the Sun and in \citet{2022bhatia_ssd} and \citet{2024Riva_ssd} for the K0V and M2V type stars in our sample. This indicates a spatial distribution of the field consistent with small-scale magnetic loops, which is already described for the Sun in \citet{2015Kitiashvili}. 

    To evaluate the effect of the local magnetic fields in the three stars studied, we compare the appearance of the granulation, the velocities, and the mean stratifications of relevant parameters from the SSD simulations against those from the HD equivalent simulations published in \citetalias{2024paperII}. As seen in the histograms and colour-maps in Fig. \ref{fig:intensities}, the bolometric intensity $I$ is significantly different between the HD and SSD runs and the intensity maps of the SSD runs show bright points in the intergranular lanes in the location of flux concentrations. The temperature, velocity, density, and Q rates at the 2D $\tau=1$ surfaces and vertical cuts in 
    Figures \ref{fig:2D_quantities_ssd_hd_1} and \ref{fig:2D_quantities_ssd_hd_2}, show the visual similarities for the SSD and HD snapshots. These similarities are quantified in Figs. \ref{fig:granule_size_ssd_vs_hd}, \ref{fig:temp_ssd_vs_hd}, \ref{fig:temp_contr_ssd_vs_hd}, and \ref{fig:vel_stat_ssd_vs_hd_1}, where the fractional area covered by granules, temperature, temperature contrast, the ratio of RMS horizontal and RMS vertical velocity, and Mach number  
    showed no significant changes when comparing the HD and SSD runs. Likewise, Fig. \ref{fig:Q_ssd_vs_hd} and Table \ref{tab:comparison_Qs_ssd} demonstrate that the Q rates in the SSD and HD runs are very close, with smaller deviations than the uncertainties arose from the implementation of the opacity treatment. 

    As shown in Fig. \ref{fig:power_spectrum}, the turbulent magnetic field produces changes in the power spectrum of temperature (for the three stars) and vertical velocity of downflows (only for the K0V and M2V stars) in the small scales. These changes could be related to the increase of these quantities in the small-scale structures of the SSD simulations compared to the HD ones or more abrupt spatial changes. 

    Kinetic energy is transformed into magnetic energy in the three stars owing to the action of the dynamo, as can be seen by the comparison of the magnetic energy in the SSD runs and the difference in kinetic energy between the SSD and HD runs (Fig. \ref{fig:energies_ssd_vs_hd_comparison}). 
    
    The power spectrum of the kinetic energy in Fig. \ref{fig:energy_power_spectra} shows that the inertial range of turbulence can be fit by the Kolmogorov power law in the case of the three stars, and that the equipartition of the kinetic and magnetic energy is reached close to the scales of energy dissipation. This picture is in agreement with previous simulations of the Sun \citep{2014rempel_SSD, 2017khomenko}. While our results suggest that the inertial range and the scale for equipartition vary with the spectral type, owing to the dependence of these scales on the resolution and boundary conditions, future studies are needed to confirm this finding.

    Consistent with previous works, we establish that SSD operates in the G2V ($T_{\mathrm{eff}}=5780 \, \mathrm{K}$, $\log g =4.4$), K0V ($4855 \, \mathrm{K}$, $4.6$), and M2V ($3690 \, \mathrm{K}$, $4.8$) considered stars. While the effects of the local magnetic fields on the mean thermodynamic structure seem negligible, we find significant differences in the bolometric intensity.  
    It should be stressed that our simulations are limited to the solar metallicity, thus, it remains as an open question if and how SSD operates in other cases.

\begin{acknowledgements}
      This work was supported by the European Research Council through the Consolidator Grant ERC--2017--CoG--771310--PI2FA and by Spanish Ministry of Science through the grant PID2021--127487NB--I00. 
      We acknowledge support from the Agencia Estatal de Investigación (AEI) of the Ministerio de Ciencia, Innovación y Universidades (MCIU) and the European Social Fund (ESF) under grant with reference PRE2018--086567. 
      NV is funded by the European Union ERC AdG SUBSTELLAR grant agreement number 101054354. 
      The authors thankfully acknowledge RES resources provided by Barcelona Supercomputing Center in MareNostrum to the activity AECT-2020-1-0021 and the resources provided by centro de Supercomputación y Bioinnovación de la Universidad de Málaga in PICASSO to the activities AECT-2022-1-0019 and AECT-2022-2-0029. We also acknowledge the use of LaPalma supercomputer from the Instituto de Astrofísica de Canarias. 
      We thank the referee for the feedback and suggestions that helped to improve the manuscript and led to the inclusion of Fig.\ \ref{fig:intensities}. 
      This research has made use of NASA's Astrophysics Data System Bibliographic Services. 
\end{acknowledgements}

\bibliographystyle{aa}
\bibliography{references}

\begin{appendix}

\onecolumn
\section{Power spectrum calculation} \label{app:FT}

The spatial power spectra from Fig.\ \ref{fig:power_spectrum} are obtained by averaging azimuthally the 2D power spectra of all snapshots at the top of the atmosphere during the stationary phase of magneto-convection. The 2D power spectra are obtained as the square of the absolute value of the 2D Discrete Fourier Transform (computed with the Fast Fourier Transform algorithm). To compute separately the power spectrum of the vertical velocity of upflows and downflows, the surface is filled with zeros where $\varv_z>0$ to compute the power spectrum of downflows and vice versa. We use a similar procedure to compute the power spectra in Fig.\ \ref{fig:energy_power_spectra}. The only difference is in the quantity to which the power spectrum is computed and the normalization: instead of the energy itself, the 2D Discrete Fourier Transform (normalized to the total number of points) of the square root  of the energy is computed, and then, the square of the result is azimuthally summed within the bins of a grid of radial spatial frequencies and divided by the length of each bin.

\section{Correlation of magnetic field and vertical velocity} \label{app:correlations}

\begin{figure*}[b!]
        \centering
        \includegraphics[width=12cm]{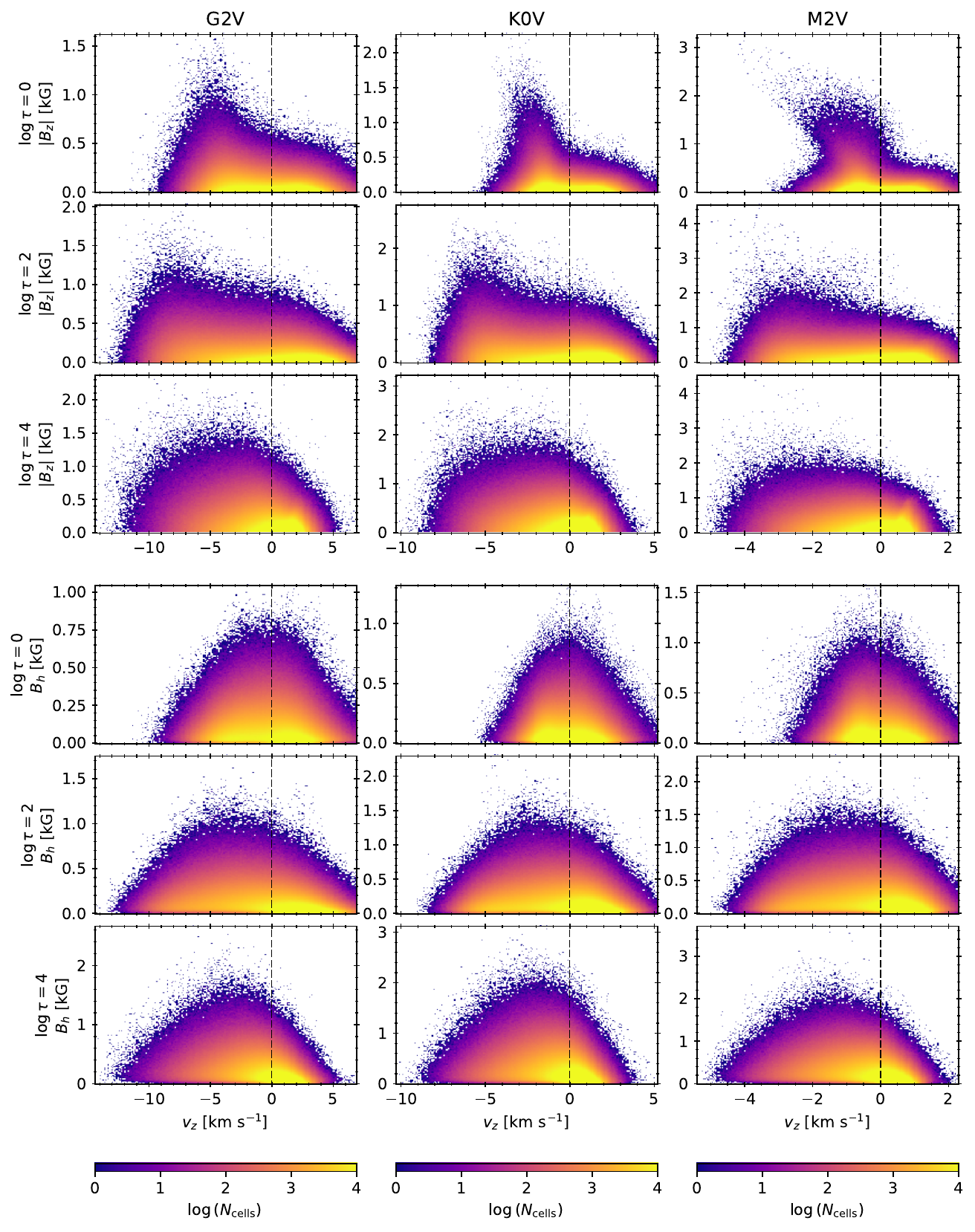}
        \caption{Logarithm of number of domain cells $N_{\mathrm{cells}}$ (colour-maps) in terms of magnetic field components $|B_z|$ (three top rows) and $B_h$ (three bottom rows) and vertical velocity $\varv_z$ in all snapshots from the SSD run of the time series of the G2V (left column), K0V (middle), and M2V (right) star. Each row shows the number of cells for a different optical depth ($\log \tau = 0, 2, 4$), indicated next to the y-axis labels at the left.}
        \label{fig:correlations}
    \end{figure*}

To understand where the magnetic fields are the largest regarding the granulation structure, Fig.\ \ref{fig:correlations} shows the number of domain cells in terms of magnetic field components and vertical velocity, for three optical depths ($\log \tau = 0, 2, 4$) and the whole time series of the SSD runs of the three stars.
The distributions of $|B_z|$ vs $\varv_z$ clearly show that the largest values of vertical magnetic field are found within the downflows. While the asymmetry in the distributions of $B_h$ vs $\varv_z$ is smaller, still the largest values of horizontal magnetic fields are in the downflows, specifically close to their edges. 

\end{appendix}

\end{document}